\documentclass[twoside,11pt]{article} 

\usepackage[left=1.25in,top=1.0in,right=1.25in,bottom=1.0in]{geometry}
\usepackage{microtype}
\usepackage{graphicx}
\usepackage{epsfig}
\usepackage{subcaption}
\usepackage{xr}
\usepackage{booktabs} 
\usepackage{amsmath,amsfonts,amsthm}
\usepackage{natbib}
\usepackage{hyperref}
\usepackage{multirow}
\usepackage{mathtools}
\usepackage{ marvosym }
\usepackage{tikz}
\usetikzlibrary{bayesnet}
\usepackage{setspace}
\usepackage{textcomp}
\usepackage{authblk}
\usepackage{bbm}
\usepackage{algorithm}
\usepackage{algpseudocode}
\usepackage{makecell}
\usepackage[capitalize]{cleveref}

\makeatletter
\newcounter{algorithmicH}
\let\oldalgorithmic\algorithmic
\renewcommand{\algorithmic}{%
  \stepcounter{algorithmicH}
  \oldalgorithmic}
\renewcommand{\theHALG@line}{ALG@line.\thealgorithmicH.\arabic{ALG@line}}
\makeatother

\newcommand{\EE}{\mathbb{E}}

\newcommand{\given}{\mid}

\providecommand\given{} 
\newcommand\SetSymbol[1][]{
  \nonscript\,#1:\nonscript\,\mathopen{}\allowbreak}
\DeclarePairedDelimiterX\Set[1]{\lbrace}{\rbrace}%
{ \renewcommand\given{\SetSymbol[]} #1 }

\newcommand{\Ind}{\mathbbm{1}}

\newcommand{\propdist}{Q_{j \given \text{-} j}}
\newcommand{\ccdist}{P(X_j \mid X_{-j})}
\newcommand{\lowerdist}{f_{j \given \text{-} j}}
\newcommand{\upperdist}{h_{j \given \text{-} j}}
\newcommand{\bootstrapdist}{B_{j \given \text{-} j}}
\newcommand{\bootstrapcdf}{C_{j \given \text{-} j}}
\newcommand{\rvknockoff}{\widetilde{X}_{j}}
\newcommand{\knockoff}{\widetilde{\mathbf{X}}_{\cdot j}}

\newcommand{\ccperm}{\mathbf{P}}
\newcommand{\ccgrid}{\mathbf{P}}

\newcommand{\bigCI}{\mathrel{\text{\scalebox{1.07}{$\perp\mkern-10mu\perp$}}}}

\newtheorem{theorem}{Theorem}

\algtext*{EndProcedure}
\algtext*{EndWhile}
\algtext*{EndIf}
\algtext*{EndFor}
\let\oldReturn\Return
\renewcommand{\Return}{\State\oldReturn}

\begin{document}

\begin{spacing}{1}
\title{The Holdout Randomization Test for Feature Selection in Black Box Models}
\date{}

\author[1]{\small Wesley Tansey\thanks{\texttt{tanseyw@mskcc.org} (corresponding author)}}
\author[3]{\small Victor Veitch}
\author[4]{\small Haoran Zhang}
\author[2]{\small Raul Rabadan}
\author[1,3,5]{\small David M.~Blei}

\affil[1]{\footnotesize Department of Epidemiology \& Biostatistics, Memorial Sloan Kettering Cancer Center, New York, NY, USA}
\affil[2]{\footnotesize Department of Systems Biology, Columbia University Medical Center, New York, NY, USA}
\affil[3]{\footnotesize Department of Statistics, Columbia University, New York, NY, USA}
\affil[4]{\footnotesize Department of Computer Science, University of Texas at Austin, Austin, TX, USA}
\affil[5]{\footnotesize Department of Computer Science, Columbia University, New York, NY, USA}
\affil[6]{\footnotesize Data Science Institute, Columbia University, New York, NY, USA}

\maketitle

\begin{abstract}%

We propose the holdout randomization test (HRT), an approach to feature selection using black box predictive models. The HRT is a specialized version of the conditional randomization test (CRT)~\citep{candes:etal:2018:panning} that uses data splitting for feasible computation. The HRT works with any predictive model and produces a valid $p$-value for each feature. To make the HRT more practical, we propose a set of extensions to maximize power and speed up computation. In simulations, these extensions lead to greater power than a competing knockoffs-based approach, without sacrificing control of the error rate. We apply the HRT to two case studies from the scientific literature where heuristics were originally used to select important features for predictive models. The results illustrate how such heuristics can be misleading relative to principled methods like the HRT. Code is available at \url{https://github.com/tansey/hrt}.

\end{abstract}

\section{Introduction}
\label{sec:introduction}
A key scientific problem is sifting through many candidate features to find explanatory signals in data. Often, predictive models are used for this task: the model is fit, error on heldout data is measured, and strong performing models are assumed to have discovered some fundamental properties of the system under study. A heuristic method (e.g. tree membership counts in random forests or coefficient magnitudes in lasso models) is then used to rank important features, with top features reported as discoveries \citep[cf.][]{barretina:etal:2012:ccle,keller:etal:2017:olfaction-prediction}. However, such heuristics provide no statistical guarantees and can result in many false positives when features are correlated. This paper presents a principled alternative to feature selection heuristics by casting the task as a multiple hypothesis testing problem.

Formally, we consider the task of selecting the subset of features ($X$) that are relevant to an outcome ($Y$). 
The inferential goal is to conduct a conditional independence test for each feature $X_{j}$, with the null hypothesis:
\begin{equation}
\label{eqn:null_hypothesis}
H_0 \colon X_{j} \bigCI Y \mid X_{-j} \, ,
\end{equation} where $X_{-j}$ is every feature in $X$ except $X_{j}$. Under $H_0$, the feature contains no information about $Y$ that is not contained in the other features. Features for which the null hypothesis is rejected are labeled as discoveries (potential drivers of the observed response).

To test the hypothesis in \cref{eqn:null_hypothesis}, \citet{candes:etal:2018:panning} proposed the conditional randomization test (CRT). The CRT works by repeatedly sampling and evaluating null samples $\rvknockoff$ that are conditionally independent of $Y$. The null samples are drawn from the \textit{complete conditional} $P(X_{j} \mid X_{-j})$, the distribution of the $j^{\text{th}}$ feature given all other features. The complete conditional serves as a valid null model for \cref{eqn:null_hypothesis} by preserving the joint dependency structure of $P(X)$ and removing any dependency between $X_{j}$ and $Y$. To evaluate each sample, the CRT compares a test statistic $T(\rvknockoff, X_{-j}, Y)$ on the null samples to the statistic on the observed sample $T(X_j, X_{-j}, Y)$. We recall the CRT in \cref{alg:crt}.

\begin{algorithm}[h]
\caption{\label{alg:crt} Conditional Randomization Test}
\begin{algorithmic}[1]
\Procedure{CRT}{features $X$, response $Y$, target $j$, test statistic $T$, null draws $K$}
\State Compute the test statistic, $t \leftarrow T(X_j, X_{-j}, Y)$.
\For {$k \leftarrow 1, \ldots, K$}
    \State Sample $\widetilde{X}_{j} \sim P(X_j \mid X_{-j})$.
    \State Compute the null statistic, $\tilde{t}^{(k)} \leftarrow T(\widetilde{X}_j, X_{-j}, Y)$.
\EndFor
\Return{A (one-sided) $p$-value,
\begin{equation*}
\hat{p}_{j} = \frac{1}{K + 1} \left( 1 + \sum_{k = 1}^{K} \mathbb{I} \left[ t \geq \tilde{t}^{(k)} \right] \right)
\end{equation*}}
\EndProcedure
\end{algorithmic}
\end{algorithm}

The vanilla CRT is statistically flexible but computationally expensive. The algorithm makes no assumptions on the form of the relationship between $Y$ and $X$. It also works with any choice of test statistic $T$. Yet for many choices of $T$, the CRT will involve substantial computation. For example, \citet{candes:etal:2018:panning} propose running the lasso and evaluating the magnitude of $\beta_j$, the coefficient for $X_j$. This requires re-running the lasso model for every null sample $\widetilde{X}_j$. In large datasets, fitting even a single model can be expensive. Refitting an expensive model thousands of times to run a CRT is prohibitive in many data analysis scenarios. 

This paper proposes an alternative: the holdout randomization test (HRT). The HRT is a specialized CRT designed to work in large datasets where it may be expensive to fit a predictive model. At a high level, the HRT works by splitting the data into train $(X, Y)$ and test $(X', Y')$ sets. A predictive model $\pi$ is fit on the training set and repeatedly evaluated on the test set. The prediction quality of $\pi(X')$ is compared to $\pi(\widetilde{X}')$, where $\widetilde{X}'$ is a copy of $X'$ that replaces the feature $X_{j}'$ to be tested with a null sample $\widetilde{X}_j'$. Intuitively, if $X_{j}'$ is predictive for $Y$ then replacing it with a null sample $\widetilde{X}_{j}'$ is likely to lead to worse predictions. A $p$-value for the hypothesis test is then approximated by repeatedly resampling $\widetilde{X}_{j}$ and comparing predictive performance under the null with performance using the original data. The HRT procedure is presented in \cref{alg:hrt}.

\begin{algorithm}[ht]
\caption{\label{alg:hrt} Holdout Randomization Test}
\begin{algorithmic}[1]
\Procedure{HRT}{training data $(X, Y)$, test data $(X',Y')$, model $\pi$, empirical risk function $\mathcal{G}$, null draws $K$}
\State Fit the model $\pi$ on the training data.
\State Compute the empirical risk on held out data, $t \leftarrow \mathcal{G}(X', Y', \pi(X'))$.
\For {$k \leftarrow 1, \ldots, K$}
    \State Sample $\widetilde{X}_{j}' \sim P(X_j \mid X_{-j})$. 
    \State Compute the empirical risk, $\tilde{t}^{(k)} \leftarrow \mathcal{G}(\widetilde{X}', Y', \pi(\widetilde{X}'))$.
\EndFor
\Return{A (one-sided) $p$-value,
\begin{equation*}
\hat{p}_{j} = \frac{1}{K + 1} \left( 1 + \sum_{k = 1}^{K} \mathbb{I} \left[ t \geq \tilde{t}^{(k)} \right] \right)
\end{equation*}}
\EndProcedure
\end{algorithmic}
\end{algorithm}

Empirical risk on held out data is the HRT test statistic. Splitting the data into train and test sets decouples model fitting from the hypothesis testing procedure. From the CRT perspective, the model is fixed and only evaluated on the test data. By decoupling model fitting and testing, the HRT makes the CRT computationally feasible for a larger class of predictive models and data analysis tasks.

The empirical risk test statistic makes the HRT applicable to any predictive model. There is no need to provide a model-specific feature importance heuristic like coefficient magnitudes or decision tree inclusion statistics. The HRT only needs to query the model for predictions in order to perform feature selection. This decoupling means the scientist does not need knowledge of the internals of the predictive model. It suffices to be able to fit a model and use it to make predictions. Being model agnostic facilitates adoption of the HRT into scientific analysis pipelines that use black box machine learning models. 

The remainder of the paper addresses several pragmatic questions regarding the HRT:
\begin{itemize}
    \item Is it possible to use the entire dataset, rather than splitting it into train and test, so as to maximize power? (\cref{subsec:extensions:cvhrt-valid,subsec:extensions:cvhrt-approx})
    \item Does the complete conditional need to be known or can we know it only proportionally? (\cref{subsec:extensions:hpt})
    \item Can we speed up the algorithm further? (\cref{subsec:extensions:hgt})
    \item How does model predictive performance relate to feature selection power? (\cref{subsec:benchmarks:model_select_and_knockoffs})
    \item If we estimate the conditional $P(X_j \mid X_{-j})$ from data, will the HRT be robust to estimation errors? If not, can we somehow err on side of caution and produce conservative $p$-values? (\cref{app:calibration})
\end{itemize}

\section{Connections to existing work}
\label{subsec:introduction:related_work}
Most classical work on feature selection relies on strong parametric
assumptions on the response function. For instance, the train-test splitting procedure of the
HRT resembles the data splitting procedure of
\citet{wasserman:roeder:2009:split-fit-select} for high-dimensional
feature selection in linear models, with the cross-validation HRT
extension (\cref{subsec:extensions:cvhrt-valid}) then mirroring the multiple
re-splitting extension of
\citet{meinshausen:etal:2009:cv-selection}. However, the HRT is
focused on the case where no parametric model is assumed. A
flurry of recent techniques have been proposed to extract information
from such black box predictive models. We relate the HRT to these
methods.

Many approaches are model-specific. Examples include iterative random forests~\citep{basu:etal:2018:iterative-random-forests}, Bayesian kernel regression~\citep{crawford:etal:2018:svm-variable-selection}, and Bayesian neural networks~\citep{liang:etal:2018:bayesian-nns}. These approaches constrain the scientist to a specific modeling choice, rather than allowing the scientist to choose the best model or build a custom one specific for their problem. The HRT allows the scientist to build the best predictive model they can, with reassurance that strong-performing models are likely to have high power.

Other methods focus on interpretation of the model, rather than testing for conditional independence. These include LIME~\citep{ribeiro:etal:2016:lime}, DeepLIFT~\citep{shrikumar:etal:2017:deeplift}, SHAP~\citep{lundberg:lee:2017:shapley}, and L2X~\citep{chen:etal:2018:l2x}. SHAP and L2X are related to the HRT in that they measure the conditional mutual information between each feature and the response. This can be seen as an optimization variant of the HRT where the statistic is change in prediction rather than prediction accuracy; however, for computational purposes both methods make simplifying independence assumptions between covariates that may lead to false positives. Even with a correct model for measuring mutual information, interpretability of the model is a related but different question. A poor model may use some true-null feature to make its predictions, in which case a correct interpretation would be to flag that variable as important to the model's prediction, even though it is conditionally independent of the true label. An incorrect model is not a problem for the HRT; it will simply have low power to detect the true signals. See~\citet{fisher:etal:2018:model-class-reliance} for an in-depth discussion on different notions of variable importance.

The leave-one-covariate-out (LOCO) inference method \citep{lei:etal:2018:loco} is a nonparametric method that assesses feature importance. LOCO inference works in the conformal inference setting \citep{vovk:2012:conditional-conformal}. Conformal inference provides coverage guarantees marginally over $X$, whereas the HRT is valid conditioned on $X=x$. Further, the LOCO test is focused on assessing the impact of a feature $X_j$ on the mean of the response $Y$. By contrast, the HRT admits any test statistic. For instance, a variable may be predictive of higher or lower variance in a response. This could be detected by the HRT but LOCO would have zero power to reject the null for this variable.

A handful of methods exist for feature selection in arbitrary black box models.

\paragraph{Conditional randomization tests (CRTs)} \citet{candes:etal:2018:panning} proposed CRTs as a generic approach for performing the conditional independence test in \cref{eqn:null_hypothesis} using any test statistic and any predictive model. A CRT repeatedly samples from the conditional null for the feature being tested and compares the test statistic under the null with the test statistic of the original data. For instance, the example test statistic in \citet{candes:etal:2018:panning} was the feature coefficient magnitude in a lasso regression. But this choice comes at a computational cost: refitting the model.  In order to calculate the lasso coefficient statistic under a null sample, the regression must be re-run using the null data. Repeating this computation many times to approximate the feature $p$-value is prohibitively expensive. The basic HRT is a special case of the CRT, where the test statistic is carefully chosen to avoid the need to refit after every null sample. 


\paragraph{Model-X knockoffs} Although
\citet{candes:etal:2018:panning} originally proposed the CRT, they discarded it in favor of the \textit{model-X
  knockoffs} approach, which avoids the computational issues outlined
above. In the model-X knockoffs framework, a ``knockoff'' feature is
generated for every feature in the dataset and the model is fit using
both the original and knockoff features. Generating the knockoffs
requires knowing (or estimating) the joint distribution over the
features; selection is performed by a model-specific variable
importance function for the original and knockoff features. The
knockoffs procedure has been extended to different feature
distributions than the original multivariate normal, such as hidden Markov models~\citep{sesia:etal:2017:hmm-knockoffs} and Gaussian mixture models~\citep{gimenez:etal:2018:gmm-knockoffs}. Machine learning approaches have also been proposed that use deep learning models to fit flexible knockoff distributions \citep{romano:etal:deep-knockoffs,jordan:etal:knockoff-gan}. Others have suggested altering certain black box models to support knockoffs, such as using a special input layer in neural networks~\citep{lu:etal:2018:deeppink-knockoffs}. While knockoffs have the appeal of computational efficiency over a generic CRT, \citet{candes:etal:2018:panning} showed in experiments that CRTs have higher power. The HRT aims to bridge the gap between the two procedures. It makes CRTs computationally tractable without sacrificing power.

\paragraph{Mimic and Classify} \citet{sen:etal:2018:mimic} propose
``Mimic and Classify,'' an approach inspired by generative adversarial
networks~\citep{goodfellow:etal:2014:gans}. Mimic and Classify fits a
conditional model of the feature, similarly to a CRT, but then also
fits a ``discriminator'' model to distinguish between samples from the
true dataset and samples from the null model, with the difference in
error magnitude as the test statistic. This approach is conceptually similar to
the HRT but differs in key ways: i) it does not leverage a pre-built
predictive model, ii) it does not provide direct $p$-value
calculation, iii) it requires fitting a discriminator per hypothesis
test, each of which may be as difficult to build as a predictive model
for the response, and iv) it implicitly ties the magnitude of the
dependency between X and Y to the test statistic, whereas the HRT is
able to detect small but consistent performance differences between
the original data and the null samples. Since Mimic and Classify does
not provide a way to control the error rate at a target level, we do
not consider it in our comparisons.

\paragraph{Conditional permutation tests (CPTs)} \citet{berrett:etal:2018:conditional-permutation} propose the CPT as a robust choice of CRT. In addition to conditioning on the the other features, the CPT also conditions on the order statistics when drawing randomizations. Thus, rather than randomly shuffling the order statistics like in a classical permutation test, each feature is shuffled in a non-uniform way. CPTs are an interesting approach and are complementary to the HRT. In particular, they are more robust than vanilla CRTs and may increase the robustness of the HRT even further. In \cref{subsec:extensions:hpt}, we merge the HRT and the CPT to create the holdout permutation test (HPT). We leverage the finite support of the empirical permutation distribution to develop a caching scheme that speeds up the MCMC algorithm from \citet{berrett:etal:2018:conditional-permutation}. In \cref{subsec:extensions:hgt}, we further extend this idea by considering combinations of discrete points--rather than permutations--to develop an HRT method that is faster than the original HRT or the HPT, while still maintaining finite-sample validity.
\newline

Finally, \citet{shah:peters:2018:conditional-independence-hardness} show that
there is no free lunch for conditional independence testing. It
is impossible to develop a test with nontrivial power
without assuming something about the joint distribution
of $(Y,X)$. We emphasize that neither the HRT nor the above methods
avoid this result. The assumption made in the HRT is that a reasonable
approximation to the complete conditional distribution of $X_j$ given
$X_{-j}$ can be
obtained. \citet{berrett:etal:2018:conditional-permutation} provide a
theoretical investigation under this assumption for CPTs.

\section{Extensions to boost power and reduce computation}
\label{sec:extensions}
The HRT provides valid $p$-values under
the assumption that we have access to the true or well-estimated complete conditional distributions.
Yet even with access to the complete conditionals, practical limitations to using the HRT arise. For instance, the availability of test data and the time to calculate the test statistics are finite. Thus, if the predictive model is computationally expensive to run and the testing dataset is small, the basic HRT may be slow and have low power to detect nonnull features. This section presents extensions to the HRT for improving the statistical and computational performance of the basic algorithm in light of these practical limitations. The empirical performance of these improvements is evaluated in \cref{sec:benchmarks}.


\subsection{The cross-validation HRT}
\label{subsec:extensions:cvhrt-valid}
Modeling pipelines typically allocate the bulk of the total data to the training set, leaving only $10$--$20\%$ for testing. Since the HRT only uses the test set and the trained model, the majority of the data will not be used for testing. In scientific settings where sample sizes are small-to-moderate, discarding this much data may result in the HRT having low power. Here we show how the train-test paradigm of the basic HRT can be extended to a cross-validation paradigm that uses the entire dataset.

\begin{algorithm}[t]
\caption{\label{alg:cvhrt-valid} Cross-Validation Holdout Randomization Test}
\begin{algorithmic}[1]
\Procedure{CV-HRT}{data split into $M$ folds: $\{(X^{(m)}, Y^{(m)})\}_{m=1}^M$, model $\pi$, empirical risk function $\mathcal{G}$, null draws $K$}
\State Initialize $t \leftarrow 0$
\For {$m \leftarrow 1, \ldots, M$}
    \State Run the fold-specific HRT, $p_j^{(m)} \leftarrow \mathrm{HRT}((X^{(-m)}, Y^{(-m)}), (X^{(m)}, Y^{(m)}), \pi, \mathcal{G}, K)$.
\EndFor
\Return{A multiplicity-corrected $p$-value,
\begin{equation}
\label{eqn:bonferroni}
\hat{p}_j = M \times \mathrm{min}(p_j^{(1)}, \ldots, p_j^{(M)}) \, .
\end{equation}
}
\EndProcedure
\end{algorithmic}
\end{algorithm}

The cross-validation holdout randomization test (CV-HRT) is presented in Algorithm~\ref{alg:cvhrt-valid}. The CV-HRT splits the data into $M$ folds, rather than the train-test split of the basic HRT. The algorithm fits $M$ models, with the $m^{\text{th}}$ model trained using the $m^{\text{th}}$ fold as the holdout set. For each model, a separate HRT $p$-value is run using the corresponding cross-validation fold as a test set; this yields $M$ fold-specific $p$-values.

Since the data are re-used in the cross-validation procedure, it is possible that there will be dependency between the fold-specific $p$-values. To correct for this possible dependency, the CV-HRT applies a Bonferroni correction to the $p$-values in \cref{eqn:bonferroni} and returns the smallest corrected value. The Bonferroni correction ensures that the resulting $p$-value from the CV-HRT retain finite-sample control of the target error rate (i.e. it will stochasticially dominate a $\mathrm{U}(0,1)$ random variable).

\subsection{Higher power via an approximate cross-validation HRT}
\label{subsec:extensions:cvhrt-approx}
The Bonferroni correction in the CV-HRT is stringent. However, it is necessary to maintain strict control over the finite-sample error rate due to the potential for dependence between the cross-validation $p$-values. In practice, we find that the dependence between the $p$-values is minimal for even small sample sizes. To boost power, we therefore propose an approximate cross-validation extension to the HRT that merges test statistics without correction.

\begin{algorithm}[t]
\caption{\label{alg:cvhrt-approx} Approximate Cross-Validation Holdout Randomization Test}
\begin{algorithmic}[1]
\Procedure{CV-HRT}{data split into $M$ folds: $\{(X^{(m)}, Y^{(m)})\}_{m=1}^M$, model $\pi$, empirical risk function $\mathcal{G}$, null draws $K$}
\State Initialize $t \leftarrow 0$
\For {$m \leftarrow 1, \ldots, M$}
    \State Fit $\pi^{(m)}$ using $(X^{(-m)}, Y^{(-m)})$.
    \State Add the fold empirical risk, $t \leftarrow t + \mathcal{G}(X^{(m)}, Y^{(m)}, \pi^{(m)}(X^{(m)}))$.
\EndFor
\For {$k \leftarrow 1, \ldots, K$}
    \State Sample $\widetilde{X}_{j} \sim \ccdist$.
    \State Initialize $\tilde{t}^{(k)} \leftarrow 0$
    \For {$m \leftarrow 1, \ldots, M$}
        \State{Add the fold empirical risk on the randomized data,
        \begin{equation*}
        \tilde{t}^{(k)} \leftarrow \tilde{t}^{(k)} + \mathcal{G}(\widetilde{X}^{(m)}, Y^{(m)}, \pi^{(m)}(\widetilde{X}^{(m)})) \, .
        \end{equation*}}
    \EndFor
\EndFor
\Return{A (one-sided) $p$-value,
\begin{equation*}
\hat{p}_{j} = \frac{1}{K + 1} \left( 1 + \sum_{k = 1}^{K} \mathbb{I} \left[ t \geq \tilde{t}^{(k)} \right] \right)
\end{equation*}}
\EndProcedure
\end{algorithmic}
\end{algorithm}

The approximate cross-validation holdout randomization test is presented in Algorithm~\ref{alg:cvhrt-approx}. As in the valid CV-HRT, the approximate CV-HRT splits the data into $M$ folds, fits $M$ models, and uses the $m^{\text{th}}$ model to make predictions for samples in the $m^{\text{th}}$ fold. Instead of calculating fold-specific $p$-values and merging them, the approximate CV-HRT calculates the average empirical risk across all folds as the test statistic. This leads to a higher-power, lower-variance estimate of the usefulness of the $j^{\text{th}}$ feature in predicting $Y$.

The theoretical validity of the approximate CV-HRT in \cref{alg:cvhrt-approx} is less clear than the provably-valid CV-HRT in \cref{alg:cvhrt-valid}. Both methods take full advantage of the dataset by building multiple models, with one valid test model per sample. The combination of these models in the case of the approximate CV-HRT, however, produces dependency between each model's predictions under the null. In simulation and in the benchmarks in \cref{sec:benchmarks}, \cref{alg:cvhrt-approx} produces valid or even conservative $p$-values. Further, because the approximate CV-HRT does not require the Bonferroni correction, the power of the procedure is increased.

Stronger predictive models in any HRT procedure are likely to capture more of the dependency structure between $X$ and $Y$. Consequently, more folds leads to more training data per model, which in turn should lead to better predictive performance per model. For the valid CV-HRT, the increase in folds comes with the tradeoff of a larger multiplicity correction term. For the approximate CV-HRT, leave-one-out cross-validation will maximize power, but this comes at the computational expense of fitting $n^*$ models, which may be prohibitive. In practice, we observe $M=5$ folds is sufficient to achieve high power in both algorithms.



\begin{algorithm}[t]
\caption{\label{alg:hpt} Holdout Permutation Test}
\begin{algorithmic}[1]
\Procedure{HPT}{training data $(X, Y)$, test data $(X', Y')=\{(X_i', Y_i')\}_{i=1}^{n'}$, model $\pi$, empirical risk function $\mathcal{G}$, null draws $K$, MCMC steps $C$}
\State Fit the model $\pi$ on the training data.
\State Initialize $n' \times n'$ matrices, $\ccperm$ and $\mathbf{T}$.
\State Compute the empirical risk on held out data, $t \leftarrow \mathcal{G}(X', Y', \pi(X'))$.
\For {$i \leftarrow 1, \ldots, n'$}
    \For {$s \leftarrow 1, \ldots, n'$}
        \State Let $\widetilde{X}_i'$ be the sample $X_i'$ with $X_{ij}'$ replaced with $X_{sj}'$.
        \State Compute the sample probability, $\ccperm_{is} \leftarrow P(X_j = \widetilde{X}_{ij}' \mid X_{-j} = \widetilde{X}_{i,-j}')$.
        \State Compute empirical risk, $\mathbf{T}_{is} \leftarrow \mathcal{G}(\widetilde{X}_i', Y_i', \pi(\widetilde{X}_i'))$.
    \EndFor
\EndFor
\For {$k \leftarrow 1, \ldots, K$}
    \State Sample permutation $Z$ via $C$ MCMC steps \citep[][Algorithm 1]{berrett:etal:2018:conditional-permutation} using cached $\ccperm$.
    \State Compute the empirical risk, $\tilde{t}^{(k)} \leftarrow \frac{1}{{n'}}\sum_{i=1}^{n'} \mathbf{T}_{i, \mathrm{rank}(Z_i)}$.
\EndFor
\Return{A (one-sided) $p$-value,
\begin{equation*}
\hat{p}_{j} = \frac{1}{K + 1} \left( 1 + \sum_{k = 1}^{K} \mathbb{I} \left[ t \geq \tilde{t}^{(k)} \right] \right)
\end{equation*}}
\EndProcedure
\end{algorithmic}
\end{algorithm}

\subsection{The holdout permutation test}
\label{subsec:extensions:hpt}
The conditional permutation test (CPT) of \citet{berrett:etal:2018:conditional-permutation} is an alternative to the conditional randomization test. The two tests are similar, but the CPT considers the permutation distribution of the feature $X_j$, rather than the total support of the independent complete conditionals. The CPT offers some benefits over the CRT, namely in terms of robustness to misspecification of the complete conditionals. However, the CPT requires a costly Markov chain Monte Carlo (MCMC) algorithm to sample over the conditional permutation distribution \citep[][Algorithm 1]{berrett:etal:2018:conditional-permutation}. The CPT algorithm may therefore be computationally prohibitive when the test set size $n'$ is large, the complete conditionals are expensive to evaluate, or the number of null samples required is large. Here we show how to adapt the HRT to the conditional permutation test to improve its scalability for these cases while still retaining the finite-sample validity of the CPT.

\cref{alg:hpt} presents the holdout permutation test (HPT). The algorithm works similarly to the basic HRT, but leverages the finite support of the permutation distribution to improve scalability. For each sample in the test set, the null distribution only contains at most $n'$ unique values.
\footnote{We assume the complete conditional distributions are continuous. If they are not, additional computational speedups can be gained. Validity of the procedure for discrete distributions still holds under Theorem 1 of \citet{berrett:etal:2018:conditional-permutation}.}
The HPT caches the probability of each unique value and the empirical risk of the predictive model under that value (Lines $4$--$9$). This requires $\mathcal{O}(n'^2)$ operations. Once cached, the sampling of new values and computing null test statistics can be done using only the cached values. Thus, when the number of null samples $K$ and MCMC samples $M$ are large, such that $KC > (n')^2$, the caching scheme in the HPT will be faster than the naive algorithm. \citet{candes:etal:2018:panning} show that $K$ must be on the order of $1/\tau$, where $\tau=qR/p$ is the BHq cutoff for $R$ rejected $p$-values. In high-dimensional feature selection problems, this number often exceeds $(n')^2$, suggesting the caching of the HPT will be useful when applying the CPT large feature sets.

\begin{algorithm}[t]
\caption{\label{alg:hgt} Holdout Grid Test}
\begin{algorithmic}[1]
\Procedure{HGT}{training data $(X, Y)$, test data $(X', Y')=\{(X_i', Y_i')\}_{i=1}^{n'}$, model $\pi$, empirical risk function $\mathcal{G}$, null draws $K$, grid size $S$}
\State Fit the model $\pi$ on the training data.
\State Compute the empirical risk on held out data, $t \leftarrow \mathcal{G}(X', Y', \pi(X'))$.
\State Initialize $n' \times (S+1)$ matrices, $\ccgrid$ and $\mathbf{T}$.
\For {$i \leftarrow 1, \ldots, n'$}
    \State Set the first column using the observed data,\[
    \ccgrid_{i1} \leftarrow P(X_j=X_{ij}' \mid X_{-j} = X_{i,-j}')\, , \qquad \mathbf{T}_{i1} \leftarrow \mathcal{G}(X_i', y_i', \pi(X_i')) \, .\]
    \State Construct an $S$ grid, $Z = (Z_1, \ldots, Z_S)$, to approximate $P(X_j \mid X_{-j} = X_{i,-j}')$.
    \For {$s \leftarrow 1, \ldots, S$}
        \State Let $\widetilde{X}_i'$ be the sample $X_i'$ with $X_{ij}'$ replaced with grid point $Z_s$.
        \State Compute sample probability and empirical risk,
        \[\ccgrid_{i,s+1} \leftarrow P(X_j=Z_s \mid X_{-j} = X_{i,-j}'))\, , \qquad \mathbf{T}_{i,s+1} \leftarrow g(\widetilde{X}_i', Y_i', \pi(\widetilde{X}_i')).\]
    \EndFor
\EndFor
\For {$k \leftarrow 1, \ldots, K$}
    \State Sample column indices $\{V_i\}_{i=1}^{n'}$ proportional to their weight in $\ccgrid_{i}$.
    \State Compute the empirical risk, $\tilde{t}^{(k)} \leftarrow \frac{1}{{n'}}\sum_{i=1}^{n'} \mathbf{T}_{i, V_i}$.
\EndFor
\Return{A (one-sided) $p$-value,
\begin{equation*}
\hat{p}_{j} = \frac{1}{K + 1} \left( 1 + \sum_{k = 1}^{K} \mathbb{I} \left[ t \geq \tilde{t}^{(k)} \right] \right)
\end{equation*}}
\EndProcedure
\end{algorithmic}
\end{algorithm}


\subsection{Faster computation via the holdout grid test}
\label{subsec:extensions:hgt}
The HRT relies on random sampling to approximate the expectation of the indicator function in \cref{alg:hrt}. Unlike other randomization tests, however, the HRT assumes a fixed model and a test statistic that is a summation of independent components. These two properties can be leveraged to substantially speed up the computation of the $p$-value approximation, in a manner similar to the caching strategy of the HPT. 

\cref{alg:hgt} presents the holdout grid test (HGT). The HGT first creates a finite grid approximation $\ccgrid$ to the complete conditional for each sample in the test dataset. It then queries the model at each point to create a matrix $\mathbf{T}$ of cached risk scores. In the test statistic computation loop (Lines $11$--$13$), the cached scores are sampled proportional to their likelihood in the complete conditional. The HGT then averages these cached scores rather than querying the predictive model. This lowers the computational complexity from $\mathcal{O}(Kn')$ to $\mathcal{O}(Sn')$. In the case of discrete distributions, the discretization step is unnecessary; the HGT would still benefit computationally from the caching strategy, however. As with the HPT, when $K$ is large (e.g. in the case of many candidate features being tested) the HGT is $1$--$2$ orders of magnitude faster than the basic HRT.

The HGT is a valid extension of the HRT. The following theorem states this formally,
\begin{theorem}
\label{thm:hgt}
The $p$-values generated by \cref{alg:hgt} are super-uniform if $H_0 \colon Y \bigCI X_j \mid X_{-j}$ is true.
\end{theorem}
We can view the grid caching scheme as a coarsening of $X_i'$ for each sample $i$. Define a new random variable $W_i$ whose support is over the discrete set of values in the $i^{\text{th}}$ grid. The local null hypothesis then becomes
\[
\tilde{H}^{(i)}_0 \colon Y \bigCI W_i \mid X_{-j} = X_{i,-j}' \, ,
\]
and \cref{alg:hgt} generates a $p$-value for the intersection null hypothesis,
\[
\tilde{H}_0 \colon \bigcap_{i=1}^{n'} \tilde{H}^{(i)}.  
\]
Testing $\tilde{H}_0$ is done by a CRT on $W = (W_1, ..., W_{n'})$. The complete conditional distribution $P(W \mid X_{-j} = X_{-j}')$ is simply the product of the individual $W_i$ complete conditionals. Since the variable $W$ is constructed without ever looking at $Y$, $\tilde{H}_0$ is true by construction if the original null $H_0$ is true. The CRT in \cref{alg:hgt} is then the HRT applied to $W$. \hfill \ensuremath{\Box}

Note that while $\tilde{H}_0$ is false only if $H_0$ is false, the converse is not true. That is, it is possible that $H_0$ is false but $\tilde{H}_0$ is true. This will be the case if $Y$ depends on $X_j$ only for some subset of values not contained in the grids. Thus in general \cref{alg:hgt} will be less powerful than the basic HRT. Simulations in \cref{subsec:benchmarks:known} investigate the difference in power between the HGT and HRT.

\section{Benchmarks}
\label{sec:benchmarks}
We investigate the basic HRT and the proposed HRT extensions in two simulation settings. First, we consider the case where the complete conditionals are known. In this setting, we show that the cross-validation HRT methods boost power and the holdout grid test speeds up computation. We then consider the case where the conditionals are unknown and must be estimated from data. Here, we show the HRT--in combination with the calibration heuristic described in \cref{app:calibration}--is robust to misspecification and finite-sample estimation error. Finally, we investigate the choice of predictive model and show that the power of the HRT is directly related to the generalization error of the predictive model. Further, with the same predictive model choice (the lasso), the HRT outperforms the model-X knockoffs approach of \citet{candes:etal:2018:panning}.

\subsection{The HRT and extensions when conditionals are known}
\label{subsec:benchmarks:known}
We benchmark the different holdout testing algorithms on a synthetic dataset sampled from a nonlinear factor model with a nonlinear response,
\begin{equation}
\label{eq:benchmark_known_generative}
\begin{aligned}
z_{ik} &\sim \mathrm{Gamma}(1, 1) \\
w_{jk} &\sim \mathcal{N}\left(0, \frac{1}{K} \right) \\
x_{ij} &\sim \mathcal{N}\left(\mathbf{z}_i^\top\mathbf{w}_j, \sigma^2 \right) \\
y_i &= \beta_1 \mathrm{tanh}(x_{i1}) + 5\mathrm{tanh}(\beta_2 x_{i2} + \beta_3 x_{i3}) + \epsilon_i \, .
\end{aligned}
\end{equation}
We set $\sigma=1$, use $K=5$ latent factors, and $\epsilon_i \sim \mathcal{N}(0,1)$. For each method, we provide the ground truth latent factors $\mathbf{Z}$ and $\mathbf{W}$; each complete conditional is independent of the other covariates conditioned on the latent factors.

We run $100$ independent trials with $3$ nonnull features and $3$ null features. The number of samples $n^*$ is varied from $20$ to $1000$. For each trial, we run the basic HRT, valid and approximate cross-validation HRT, HPT, and HGT. The HPT uses $M=50$ MCMC steps per null sample and the HGT uses a grid size of $S=50$. Each method samples $K=10000$ null samples to estimate the feature $p$-value. After $p$-value calculation, we select features using Benjamini-Hochberg (BH) \citep{benjamini:hochberg:1995:bh} with nominal FDR level of $10\%$.

We analyze four predictive model choices: ordinary least squares, lasso with regularization parameter chosen via cross-validation, Bayesian ridge regression with automatic relevance determination, and random forests (RFs) with $20$ trees. Results for all four models are similar. We focus here on RFs; all results for all predictive models are presented in \cref{app:benchmarks_known}.

\begin{table}[t!]
\centering
\begin{tabular}{lccccccc}
\toprule
Algorithm & Splits                    & $n=20$    & $n=50$    & $n=100$  & $n=200$   & $n=500$   & $n=1000$  \\
\midrule
HRT (\cref{alg:hrt}) & 1        & 0.05 & 0.18 & 0.38  & 0.64  & 0.7   & 0.7    \\
\midrule
\multirow{6}{*}{\makecell{CV-HRT \\ (\cref{alg:cvhrt-valid})}} & 2  & 0.08 & 0.2  & 0.5   & 0.67  & 0.71  & 0.7    \\
& 3  &  0.06 & 0.19 & 0.42  & 0.63  & 0.7   & 0.68   \\
& 4  &  0.06 & 0.15 & 0.38  & 0.63  & 0.71  & 0.69   \\
& 5  &  0.03 & 0.13 & 0.35  & 0.58  & 0.7   & 0.71   \\
& 10 &  0    & 0.07 & 0.19  & 0.46  & 0.67  & 0.69   \\
& 20 &  0    & 0.02 & 0.13  & 0.26  & 0.61  & 0.69   \\
\midrule
\multirow{6}{*}{\makecell{CV-HRT \\ (\cref{alg:cvhrt-approx})}} & 2 & 0.1  & 0.29 & 0.51  & 0.68  & 0.7   & 0.7    \\
& 3  & 0.09 & 0.27 & 0.56  & 0.7   & 0.69  & 0.7    \\
& 4  & 0.14 & 0.29 & 0.58  & 0.7   & 0.72  & 0.69   \\
& 5  & 0.1  & 0.31 & 0.54  & 0.69  & 0.71  & 0.71   \\
& 10 & 0.12 & 0.34 & 0.56  & 0.68  & 0.71  & 0.71   \\
& 20 & 0.08 & 0.29 & 0.56  & 0.7   & 0.69  & 0.7    \\ 
\bottomrule
\end{tabular}
\caption{\label{tab:benchmark_cvhrt} True positive rate for the random forest predictive model in \cref{subsec:benchmarks:known}. Both cross-validation methods improve on the basic HRT in terms of sample efficiency. The approximate CV-HRT method in \cref{alg:cvhrt-approx} has the highest power as it does not use the multiple testing correction of \cref{alg:cvhrt-valid}.}
\end{table}


\paragraph{Cross-validation extensions.} \Cref{tab:benchmark_cvhrt} presents the results for the two cross-validation HRT methods. As the sample size approaches $n=1000$, the basic HRT performs as well as both cross-validation variants. When the sample size is smaller, both CV-HRT methods can outperform the basic method. For the CV-HRT in \cref{alg:cvhrt-valid}, a smaller number of splits leads to higher power since the multiple testing correction term is smaller. The approximate CV-HRT in \cref{alg:cvhrt-approx}, in contrast, does not account for dependence and consequently generally benefits from more splits but sees diminishing returns beyond $4$ folds. Despite being only an approximate method, \cref{alg:cvhrt-approx} produces empirically valid or even conservative null test statistics and controls the FDR across all simulation settings; see \cref{app:benchmarks_known} for detailed results.

\begin{table}[t!]
\centering
\begin{tabular}{cccccccc}
\toprule
& Algorithm      & $n=20$    & $n=50$    & $n=100$  & $n=200$   & $n=500$   & $n=1000$  \\
\midrule
\multirow{3}{*}{TPR} &  HRT (\cref{alg:hrt})  & 0.05 & 0.18 & 0.38  & 0.64  & 0.7   & 0.7    \\
 &  HPT (\cref{alg:hpt})  & 0.07 & 0.19 & 0.39  & 0.63  & 0.71  & 0.71   \\
 &  HGT (\cref{alg:hgt}) & 0.04 & 0.18 & 0.34  & 0.61  & 0.7   & 0.69   \\
\midrule
\multirow{3}{*}{Time} & HRT (\cref{alg:hrt}) & 5.88  & 5.99  & 6.29 & 6.94  & 9.55  & 14.02 \\
 & HPT (\cref{alg:hpt}) & 13.16 & 13.72 & 14.04 & 15.50 & 19.75 & 26.39 \\
 & HGT (\cref{alg:hgt}) & 0.02  & 0.03  & 0.04  & 0.06  & 0.13  & 0.25   \\
\bottomrule
\end{tabular}
\caption{\label{tab:benchmark_timing} Power and timing (in seconds) results for the HRT, HPT, and HGT using random forests as the predictive model.}
\end{table}

\paragraph{Permutation and grid extensions.} \Cref{tab:benchmark_timing} presents the results for the two alternative holdout testing methods. The HGT runs orders of magnitude faster than both the HRT and HPT. This comes at no cost in power relative to the HRT; the HPT has power benefits with small sample sizes, but comes at a higher computational cost. As with the CV-HRT, all three methods control the FDR across all simulations.

\subsection{Performance when conditionals are unknown}
\label{subsec:benchmarks:unknown}
To evaluate a more practical scenario, we next study the HRT on a variant of a benchmark from a recent paper on feature selection in Bayesian neural networks \citep{liang:etal:2018:bayesian-nns}. In this scenario, we do not assume the complete conditionals are known and instead estimate them from the observed data. We generate $100$ independent datasets of $500$ samples each with a nonlinear ground truth regression model,
\begin{equation}
\label{eqn:benchmark_ground_truth}
y = \sum_{j=0}^{9} \left[ w_{4j} x_{4j} + w_{4j+1} x_{4j+1} + \text{tanh}(w_{4j+2}x_{4j+2} + w_{4j+3}x_{4j+3})\right] + \sigma\epsilon \, ,
\end{equation}
where $\sigma = 0.5$ and $\epsilon \sim \mathcal{N}(0,1)$. The $500$ features are generated to have $0.5$ correlation coefficient with each other,
\begin{equation}
\label{eqn:benchmark_covariates}
x_j = (\rho + z_j) / 2\, , \quad j = 1, \ldots, 500 \, ,
\end{equation}
where $\rho$, $z_j$, and $w_j$ are independently generated from $\mathcal{N}(0,1)$.

The ground truth in \eqref{eqn:benchmark_ground_truth} uses the first $40$ features, representing true signals; each sample also contains $460$ null features. We note that \citet{liang:etal:2018:bayesian-nns} only had $4$ signal features in their simulation. In preliminary simulations, the HRT had nearly $100\%$ power and $0\%$ FDP in this setting; we extend the experiment to $40$ features to make the benchmark more challenging and model comparisons more informative.

We choose this setup for several reasons: (i) it is a nonlinear ground truth, (ii) most features are not involved in the response, (iii) all features are highly correlated, and (iv) it is compatible with the available implementation of model-X knockoffs~\citep{candes:etal:2018:panning} for the lasso; we compare to this implementation in \cref{subsec:benchmarks:model_select_and_knockoffs}. 

For all experiments, we use a mixture density network~\citep{bishop:1994:mixture-density-networks} with $5$ components as the conditional density estimator in the HRT. Note that this flexible model is effectively misspecified and, at low sample sizes such as this experiment, is highly variable in its predictions across random restarts. We use this as it represents a realistic scenario where conditional densities are likely to be poorly estimated.

For each feature, we fit a bag of $b=100$ bootstrap estimators and use the data-adaptive calibration technique in \cref{app:one_way_ks} to choose the approximate lower and upper bounds. We use mean-squared error (MSE) as the empirical risk function. We select significant features with a target FDR of $10\%$, using the Benjamini-Hochberg procedure (BH) \citep{benjamini:hochberg:1995:bh} for multiple testing correction. For BH correction to be theoretically valid, the $p$-values must satisfy the \textit{positive regression dependence on a subset} (PRDS) criterion~\citep{benjamini:yekutieli:2001:dependence}. The HRT provides no guarantees about the dependence between the $p$-values. To fully control FDR under dependence, \citet{benjamini:yekutieli:2001:dependence} provide a more conservative procedure (typically called BY correction). We present results using BH selection since it is the most common procedure and empirically robust; results using BY correction are provided in \cref{app:by_benchmarks}. 
 
\begin{figure}[t]
\centering
\includegraphics[width=0.95\textwidth]{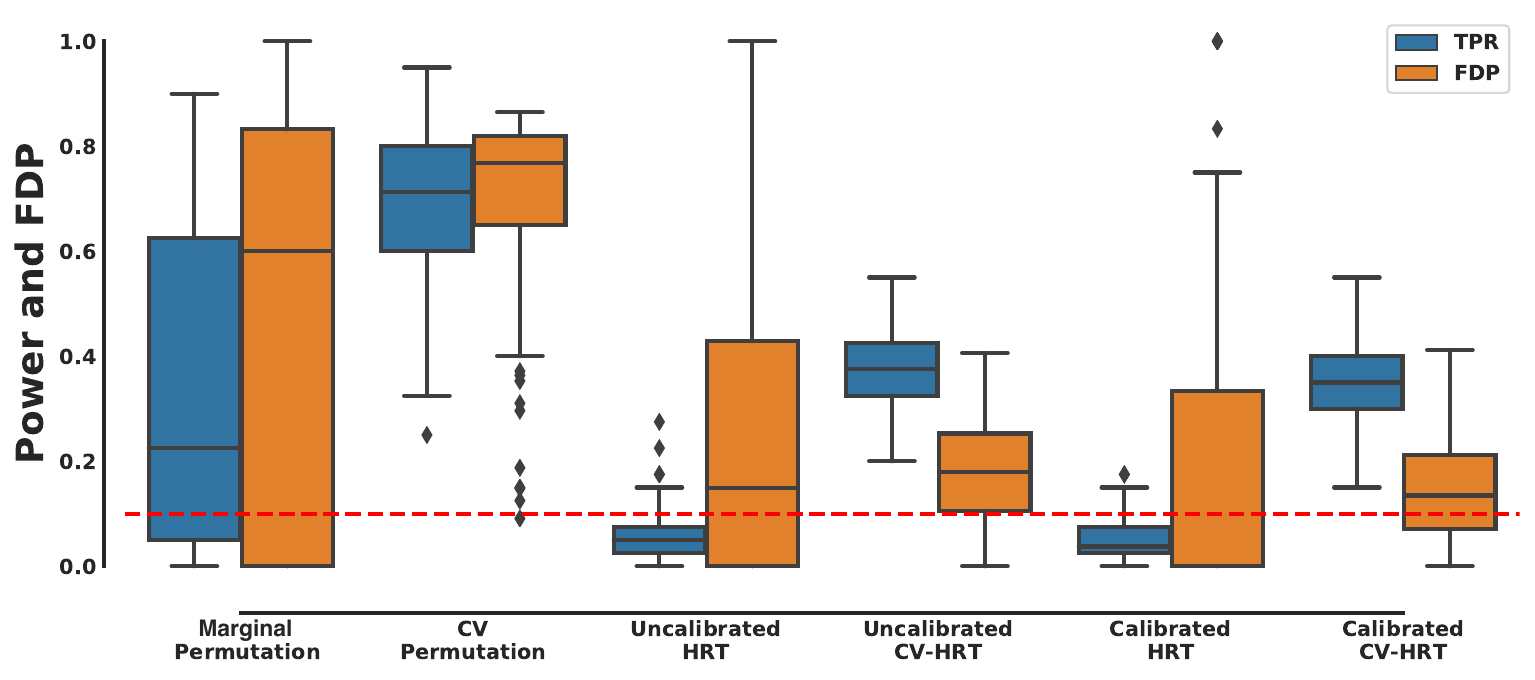} 
\caption{\label{fig:benchmark_results} Power and FDP results for each HRT variant on the benchmark simulation. The calibration technique from \cref{app:calibration} adjusts the sample weights to achieve tighter control over the FDR at the specified $10\%$ level (dashed red line). Using cross-validation for predictive modeling increases power and reduces variance by using the entire dataset rather than just a held out subset.}
\end{figure}

We first compare several variants of the HRT procedure to investigate the effect of each component. For each dataset, the predictive model is a neural network with $2$ hidden layers of $200$ nodes each and ReLU activation functions; we fit the model using RMSprop with fixed learning rate $3\times10^{-5}$. We use an $80\%/20\%$ train/test split for the basic HRT and $5$-fold cross-validation for the CV-HRT in \cref{alg:cvhrt-approx}. \cref{fig:benchmark_results} shows the power and FDP results for the different variants. We discuss the individual variants and results below.\footnote{Since the FDP results have high variance, the box plots show the median rather than the mean. The mean false discovery proportions for the uncalibrated and calibrated CV-HRT are $18.46\%$ and $14.48\%$, respectively.}

\paragraph{Marginal Permutation.} A common approach to testing for feature importance in many applications is the classical permutation test, where the values of a given feature are shuffled. Combining this with empirical risk on held out data was suggested by \citet{breiman:2001:two-cultures} as a feature selection technique. To be valid, permutation testing requires that features are independent, which is explicitly not the case in the benchmark nor in many---if not most---real world datasets. The two left-most results in \cref{fig:benchmark_results} show performance using a permutation approach, rather than the complete conditional, in both the basic and cross-validation HRT. The misspecified independence assumption causes an extreme inflation in the empirical false discovery rate to nearly $80\%$ instead of the target $10\%$.

\paragraph{Calibration.} The middle two results in \cref{fig:benchmark_results} show performance using a single, uncalibrated estimation of $\ccdist$ in the basic and cross-validation HRT. The estimated conditional is a much better approximation of the true feature distribution than the marginal permutation above, but is still insufficient to control the empirical FDR at the target level. By contrast, the two right-most results show that the calibrated model better conserves FDR, with minimal impact on power compared to the uncalibrated model. Additional investigation of the effects of calibration are available in \cref{app:benchmarks_unknown}.

\paragraph{Basic HRT vs. CV-HRT.} The right-most results in \cref{fig:benchmark_results} demonstrate the effect of using a basic train-test split HRT versus a 5-fold cross-validation HRT. Cross-validation has the effect of boosting power and reducing variance compared to the basic HRT.

\subsection{Model selection and knockoffs comparison}
\label{subsec:benchmarks:model_select_and_knockoffs}
We next compare several choices of predictive model on the same benchmark task as above. Specifically, we ran the CV-HRT procedure with the neural network model from the previous section and the following alternative predictive models:
\begin{itemize}
    \item Ordinary least squares (OLS)
    \item Partial least squares (PLS) with $10$ components.
    \item Lasso with penalty parameter chosen via $5$-fold cross-validation on each training set.
    \item Elastic net with both penalty parameters chosen via $5$-fold cross-validation on each training set.
    \item Bayesian ridge with normal-inverse-gamma priors for coefficients and weakly informative scale hyperpriors.
    \item Kernel ridge with cubic polynomial kernel with a fixed penalty weight of $1$.
    \item Support vector with radial basis function (RBF) kernel.
    \item Random forest with $20$ trees.
\end{itemize}
All model hyperparameters not specified were set to their default in the \texttt{scikit-learn} Python package. We also compare to a lasso knockoff model using coefficient difference statistic; we use the implementation provided in the \texttt{knockoffs} R package.

\begin{figure}[th!]
\centering
\begin{subfigure}{0.95\textwidth}\includegraphics[width=\textwidth]{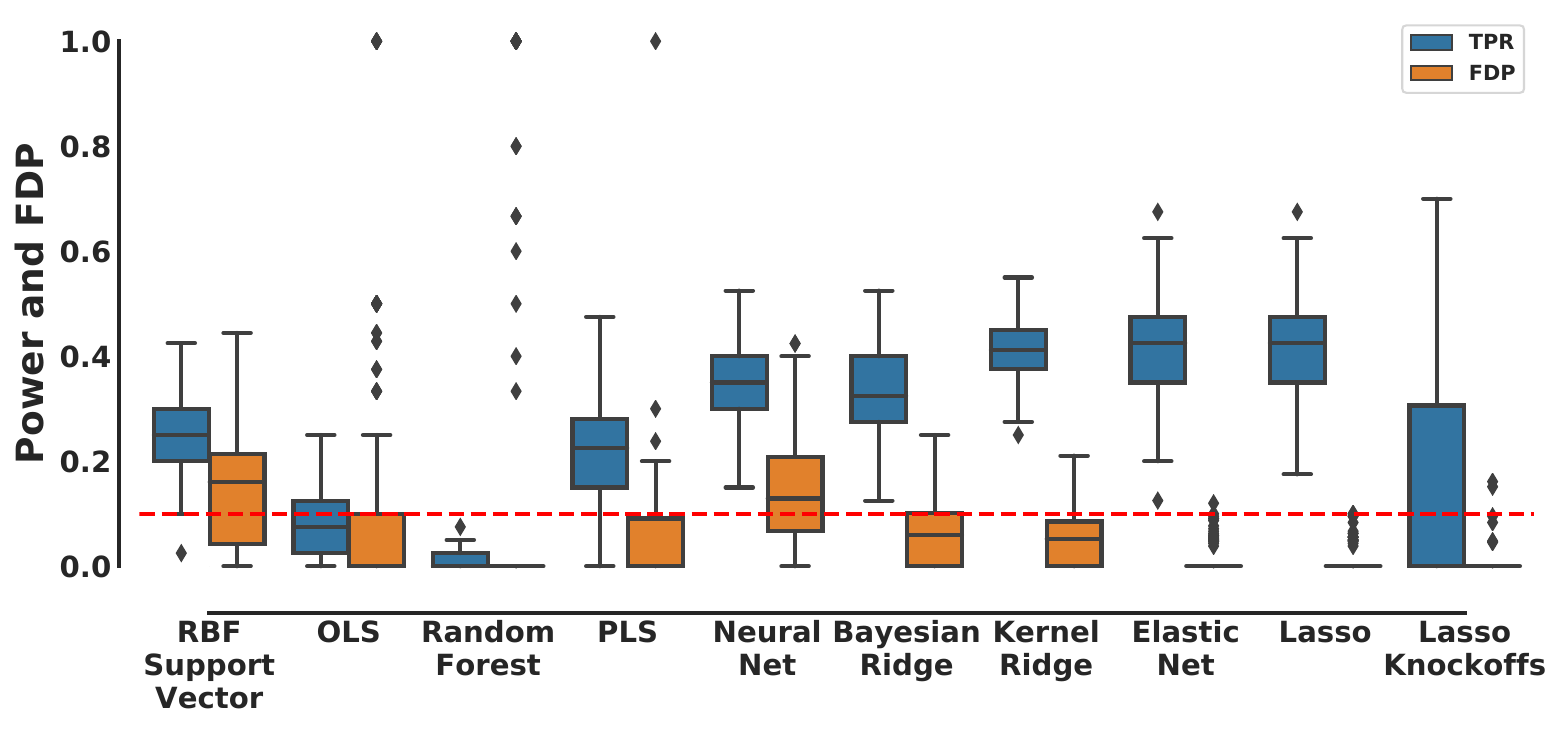}\end{subfigure}
\begin{subfigure}{0.45\textwidth}\includegraphics[width=\textwidth]{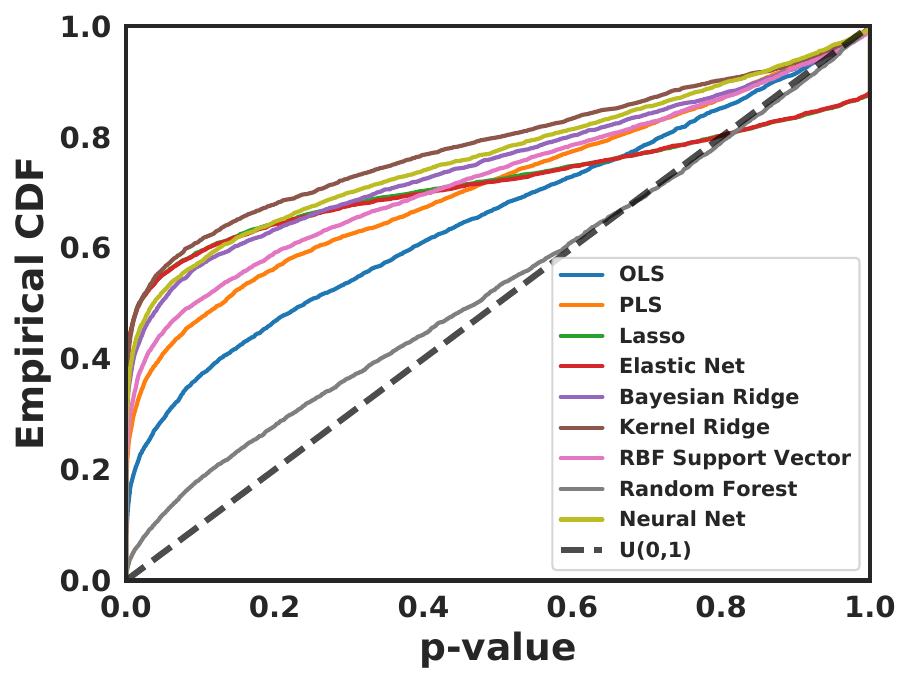}\end{subfigure}
\begin{subfigure}{0.45\textwidth}\includegraphics[width=\textwidth]{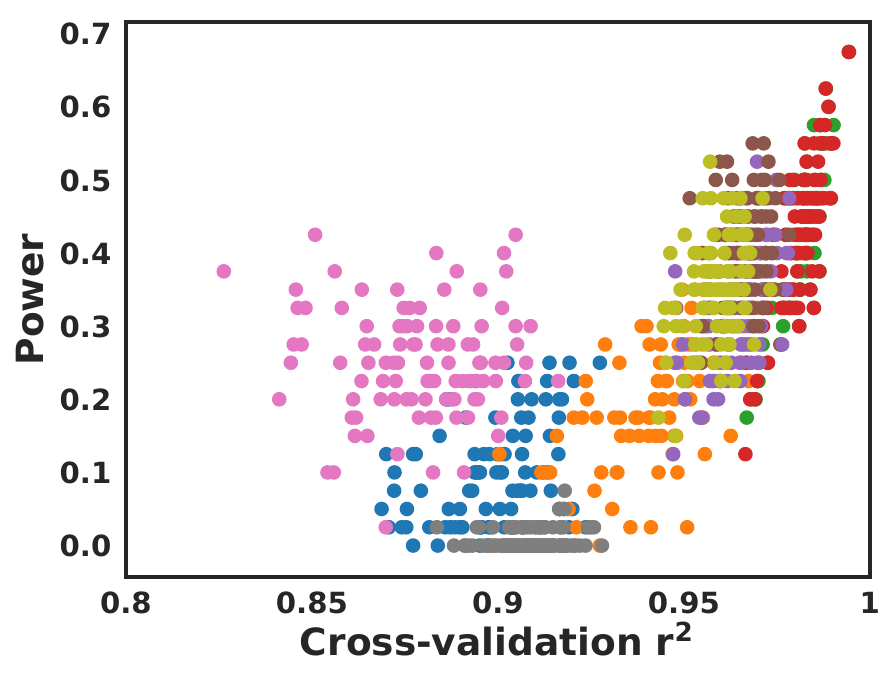}\end{subfigure}
\caption{\label{fig:predictors} Top: Power and FDP for different choices of predictive model used in the CV-HRT, ordered by empirical risk of the predictive model. The right-most result is for a lasso model using model-X knockoffs for feature selection. Bottom left: distribution of p-values for signal features using each predictive model. Bottom right: predictive model performance versus feature selection power for each model and independent trial. Colors in the bottom right plot correspond to those in the bottom left.}
\end{figure}


\cref{fig:predictors} (top) shows the power and FDP results for each model. The predictive models are ordered by average cross-validation $r^2$ across all trials. CV-HRT power roughly increases with the quality of the predictive model, with the lasso model producing the highest power results. \cref{fig:predictors} (bottom right) demonstrates this with more granularity. It shows the power for each individual model on each trial as a function of the individual model $r^2$. Even within the same model class, the models that generalize better tend to have higher power in the CV-HRT.

A notable exception is the random forest model, which has low power across all trials. This may be due to the nature of the model. Random forests average predictions from an ensemble of decision trees. Each tree predicts by recursively dividing into a series of half-planes, with each half-plane based on a single feature. Most null samples in the HRT are near, but slightly different than, the original data. It is likely that a small change to the original data will not change the random forest prediction, since it may not cause the feature to cross any half-plane. Since the comparison of the test statistic under the original model with each null sample statistic is not strict in the HRT, the insensitivity of random forests to small data changes leads to lower power. One solution to this is to use more trees in the forest, with data subsampling or bootstrapping to produce many different half-planes for each feature. 

\cref{fig:predictors} also presents the results of a lasso knockoffs model using the coefficient magnitude as a test statistic, as proposed by \citet{candes:etal:2018:panning}. The power of the CV-HRT lasso model in this simulation is approximately $3$x more than knockoffs ($45\%$ versus $15\%$ on average).

\section{Case studies from the scientific literature}
\label{sec:case_studies}
We demonstrate the usefulness of the HRT on two datasets from real experiments. The first dataset measures the genomic profile of hundreds of cancer cell lines and their response to an anti-cancer drug. The second measures the molecular structure of hundreds of chemicals and their perceived olfactory properties. In both experiments, the original scientific analyses followed the same template: (i) a large number of features were gathered about the target, (ii) a predictive model was fit, (iii) a model-specific feature importance heurstic was used to rank features, and (iv) top-ranked features were reported as discoveries.

In both case studies, we followed the same procedures for steps (i) and (ii), but then replaced (iii) and (iv) with the calibrated CV-HRT (\cref{alg:cvhrt-approx}). We use the same conditional estimator as in \cref{subsec:benchmarks:known}, but for numerical stability take the first $100$ principal components as inputs rather than raw features. We briefly describe each experiment and the original analysis approach, then present a comparison of the heuristics employed with the discoveries made by the HRT.

\paragraph{Drug response experiment} The first dataset is a study of anti-cancer drug response in cancer cell lines~\citep{barretina:etal:2012:ccle}. Multiple drugs were tested against hundreds of cell lines; phenotypic response was measured as the area under the dose-response curve (AUC). Each cell line was analyzed to obtain gene mutation and expression features. The scientific goal is to discover the genomic features associated with drug response. Genomic features were first screened to filter out features with less than $0.1$ magnitude Pearson correlation to the AUC. An elastic net model was fit for each drug, with hyperparameters chosen via $10$-fold cross-validation. Features were then ranked by average coefficient magnitude. We choose a single drug, PLX4720, as an illustrative example. The results presented are similar to the original publication results for PLX4720, though not identical. We followed the analysis to the best of our ability, but the publicly available data is newer than the dataset used in the original publication. Our results are therefore different than those published, but not meaningfully so.

\paragraph{Olfactory perception experiment} The second dataset is a study of the perceived fragrant properties of various molecules~\citep{keller:etal:2017:olfaction-prediction}. Several hundred human subjects smelled and rated each molecule across $20$ different categories. For each molecule, several thousand descriptive features about its chemical structure were measured. A random forest model was fit for each category to predict the average human rating given the features for a molecule. Different types of molecular descriptors were tried, with two sets (Dragon and Morgan descriptors) being selected based on performance on a held out test set. Features were then ranked by the random forest feature importance heuristic in \texttt{scikit-learn}. This heuristic estimates the expected number of samples in which a feature is used, based on how often and how deep it appears in the constituent trees. We again choose a single illustrative example, the \textit{Bakery} category.

To increase power, we use the heuristics as a filter on which features to test in both case studies. Features with less than $10^{-3}$ heuristic importance (coefficient magnitude in elastic net and expected usage in random forests) are ignored. This leaves $873$ features for the drug response dataset and $93$ features for the olfactory perception dataset. This filtering step does not affect the statistical guarantees of the HRT since it is not based on any of the held out test data. It only avoids testing features that are completely ignored by the model and which the HRT will have no power to detect, if they are signals.

\begin{table*}
\centering
\begin{small}
\hspace{-0.8cm}\begin{subtable}{.45\textwidth}
\centering
\caption{\label{tab:ccle_elastic_net}Elastic net for cancer drug response}
\begin{tabular}{l|c|l}
\toprule
\thead{Genomic \\ Feature} & \thead{Imp. \\ score} & \thead{Est. \\ $p$-value} \\
\midrule
BRAF V600E Mut &      0.0975 & $\leq 10^{-5}$* \\
RP11-208G20.3 &          0.0715 & 0.0065* \\
RP6-149D17.1 &      0.0665 & 0.2108 \\
RNU6-104P  &      0.0652 & 0.9970 \\
RNA5SP184  &          0.0634 & 0.3359 \\
VPS13B Mut &   0.0557 & 0.0042* \\
RP11-567M16.3  &    0.0535 & 0.0079* \\
MTMR11 Mut &  0.0525 & 0.0821 \\
ZNF549  &        0.0524 & 0.1915  \\
HIP1 Mut &      0.0519 & $\leq 10^{-5}$* \\
\end{tabular}
\end{subtable}
\hspace{0.5em}
\begin{subtable}{.45\textwidth}
\centering
\caption{\label{tab:olfaction_random_forest}Random forests for olfactory perception}
\begin{tabular}{l|c|l}
\toprule
\thead{Molecular \\ Feature} & \thead{Imp. \\ score} & \thead{Est. \\ $p$-value} \\
\midrule
Isovanillin & 0.2629 & 0.0012* \\
Vanillin isobutyrate & 0.0528 & 0.2553 \\
Ethyl vanillin & 0.0481 & 0.0430 \\
Ethyl vanillin acetate & 0.0258 & 0.6066 \\
Protocatechualdehyde & 0.0236 & 0.1515 \\
Vanillin acetate & 0.0232 & 0.2845 \\
2-Formylimidazole & 0.0186 & 0.8132 \\
R7e+ & 0.0179 & 0.5806 \\
Ethyl Isovalerate & 0.0123 & 0.7837 \\
SM05 AEA(ri) & 0.0099 & 0.2751 \\
\end{tabular}
\end{subtable}
\end{small}
\caption{\label{fig:real_examples} Two examples of predictive modeling with heuristic post-hoc feature importance ranking in scientific studies. a) Genomic features were used to predict cell line response to treatment with the drug PLX-4720~\citep{barretina:etal:2012:ccle}. b) Molecular features were used to predict perceived fragrant properties of molecules~\citep{keller:etal:2017:olfaction-prediction}. Asterisks indicate features selected by the HRT at a $20\%$ FDR threshold.}
\end{table*}

\cref{fig:real_examples} shows the top $10$ ranked features in both datasets, following their respective heuristics. Beside each feature, we report its heuristic importance score and the $p$-value assigned by the HRT. The heuristic rankings correlate poorly with the (theoretically grounded) $p$-value assigned by the HRT. Features with an asterisk denote those selected by the HRT with Benjamini-Hochberg correction at a 20\% FDR threshold. It is impossible to know whether these features are all true positives, but the HRT demonstrates that the statistical evidence from the model does not match with the ranking heuristics.

\begin{table*}
\centering
\begin{tabular}{l|c|c|c}
\toprule
\thead{Genomic \\ Feature} & \thead{Heuristic \\ ranking} & \thead{Coefficient \\ magnitude} & \thead{Est. \\ $p$-value} \\
\midrule
CNTN1 Mut & 11 & 0.0512 & 0.0059 \\
RNU6-448P & 12 & 0.0486 & 0.0000 \\
MIR4482-1 & 14 & 0.0474 & 0.0004 \\
MTND4P25 & 19 & 0.0457 & 0.0010 \\
RP11-585F1.8 & 20 & 0.0451 & 0.0058 \\
RN7SL528P & 21 & 0.0433 & 0.0034 \\
KRTAP23-1 & 25 & 0.0384 & 0.0006 \\
GS1-24F4.3 & 29 & 0.0372 & 0.0000 \\
FLT3 Mut & 31 & 0.0363 & 0.0000 \\
RNU6-1287P & 32 & 0.0360 & 0.0000 \\
RP11-575H3.1 & 44 & 0.0294 & 0.0035 \\
SNAPC3 & 51 & 0.0268 & 0.0012 \\
RP11-541E12.1 & 55 & 0.0251 & 0.0000 \\
RP11-488I4.2 & 70 & 0.0225 & 0.0012 \\
RNA5SP234 & 72 & 0.0224 & 0.0000 \\
CDC42BPA Mut & 76 & 0.0213 & 0.0000 \\
NEURL & 79 & 0.0205 & 0.0000 \\
RP11-481A12.2 & 82 & 0.0203 & 0.0002 \\
HCP5 & 86 & 0.0200 & 0.0011 \\
CTC-539A10.7 & 91 & 0.0190 & 0.0062 \\
PIK3R4 Mut & 98 & 0.0183 & 0.0025 \\
SLC35G6 & 115 & 0.0165 & 0.0008 \\
PIP5K1A Mut & 122 & 0.0162 & 0.0002 \\
CCND2 Mut & 171 & 0.0123 & 0.0026 \\
CDC37L1 & 178 & 0.0121 & 0.0000 \\
CSPG4 Mut & 211 & 0.0094 & 0.0003 \\
RP5-1195D24.1 & 213 & 0.0094 & 0.0000 \\
DIP2C Mut & 256 & 0.0081 & 0.0003 \\
RIOK3 Mut & 358 & 0.0055 & 0.0049 \\
POLR2J4 & 401 & 0.0047 & 0.0031 \\
CACNA2D2 & 409 & 0.0046 & 0.0060 \\
RNA5SP280 & 420 & 0.0044 & 0.0020 \\
NTSR1 Mut & 672 & 0.0018 & 0.0002 \\
\end{tabular}
\caption{\label{tab:extra_hrt_discoveries} Additional genomic features selected by the HRT at a $20\%$ false discovery rate threshold in the drug response study, but not in the top 10 ranking of coefficient magnitude.}
\end{table*}

\paragraph{Additional discoveries by the HRT.}
\label{subsec:case_studies:hrt_discoveries}
In the drug response experiment, the HRT selected additional features. \cref{tab:extra_hrt_discoveries} shows all discoveries by the HRT at a $20\%$ FDR threshold that were not ranked in the top $10$ by the heuristic. Alongside each feature we show its $p$-value estimate, heuristic score, and heuristic ranking. Several features were found to have significant predictive power despite their estimated model effect size being relatively low. This suggests there may be new potential targets of therapy worthy of follow-up investigation.

\section{Discussion}
\label{sec:discussion}
When Leo Breiman wrote about the two cultures in statistics \citep{breiman:2001:two-cultures}, he divided statisticians into the \textit{data modelers} and the \textit{algorithmic modelers}. Data modelers relied on inflexible, idealized parametric models that make unrealistic assumptions about the true data generating distribution in order to gain the mathematical convenience of verification tools like goodness-of-fit tests. Algorithmic modelers chose to treat the latent function mapping features to response as a black box. In this latter paradigm, predictive performance on held out data was the only way to truly measure the strength of a model. Breiman estimated at the time that $98\%$ of the mainstream analyst world fell into the data modeling crowd, while a mere $2\%$ were algorithmic modelers.

Nearly $20$ years later, the tables have turned. A confluence of factors (better numerical computing tools, the rise of Computer Science as one of the most popular majors in university, the dawn of the ``Big Data'' era, high throughput screening techniques in science, and the practical successes of machine learning) has led us to a world where algorithmic modeling is the norm and test error is the gold standard for model evaluation. The gravity of algorithmic modeling has grown so strong that it has even pulled in many scientific fields, leading to prediction challenges like the DREAM series~\citep{stolovitzky:etal:2007:dream-intro}. These challenges publicly release scientific data and teams compete to build the best model based solely on performance on held out test data. One would be hard pressed to argue that the algorithmic modelers are in the minority today.
 
Yet as we have shown, this shift has come at the cost of reliably extracting understanding from the data, what Breiman called the \textit{information} goal. Articles in premier scientific journals, such as the two case studies in \cref{sec:case_studies}, now commonly use machine learning methods to build predictive models then derive scientific conclusions from ad-hoc model interpretation. These heuristic approaches rest on unstable ground and may lead scientists astray by convincing them that a strong signal exists where there is only noise. Breiman himself nearly solved this by effectively proposing to use what we call the marginal permutation HRT (\cref{subsec:benchmarks:known}) as a means for interpreting random forests, but he did not consider the need to account for feature dependency structure in his test. It is fitting then that rigorously extracting information from black box models, as the HRT does, requires using \textit{more} black box modeling (i.e. conditional density estimators) to disentangle the features and perform a reliable conditional independence test.


A key question left open by the HRT is how to estimate complete conditionals reliably. The heuristic proposal in \cref{app:calibration} works in our experiments, but it is ad hoc and requires a heavy computation to generate bootstrapped confidence intervals. The same question is raised for other methods such as model-X knockoffs, which in practice all estimate the knockoff distribution from data without guarantees on the results. Future theoretical work is needed to understand how to estimate complete conditionals with minimal assumptions and statistical guarantees on the resulting CRT. Similarly, it may be for certain models and datasets to derive an optimal splitting scheme that maximizes power in a robust class of response functions. The theoretical foundations of how to allocate training and testing splits is also left for future work. 

There are several additional directions to explore for extending the HRT to other analysis settings. In some models, such as empirical Bayes models that use predictive models for prior estimation \citep{tansey:etal:icml:2018:bbfdr,tansey:etal:2020:dose-response,tansey:etal:2020:debt}, inference is still too expensive to run the HRT. Efficiently searching over the space of potential features to test, using an approach like Bayesian optimization~\citep{shahriari:etal:2016:bayesian-optimization-review} or multi-armed bandit learning~\citep{gandy:hahn:2017:quickmmctest}, would facilitate applying the HRT to compute-intensive models. In other scenarios, such as image classification, language modeling, and time series analysis, the features themselves have a more complicated structure and selecting an individual feature is not necessarily the inferential goal. For example, a pixel location being significant is not usually an interesting discovery in image classification. A conceptual layer needs to be added to extract meaningful insight in these types of models.

Finally, the HRT has the added benefit of allowing arbitrarily fine-grained questions to be asked. For instance, the scientist can partition the dataset based on some other feature, at which point the null hypothesis is $X_j \bigCI Y \mid X_{k}=x, X_{-(j,k)}$. This is a common test in biology, where predictive models are often trained on all of the data but questions are then asked about feature dependencies for specific samples, such as a specific type of cancer or class of drug. We plan to explore these data-partitioning scenarios in future work.

\paragraph{Acknowledgments.}
The authors thank Yixin Wang, Daniel Hsu, Ioan Filip, Scott Linderman, Christopher Tosh, Jackson Loper, and Lucas Janson for helpful conversations in the early stages of this work.

\end{spacing}

\begin{small}
\bibliographystyle{abbrvnat}
\bibliography{main}
\end{small}

\appendix

\section{Conservative estimation by bootstrap sample reweighting}
\label{app:calibration}
The error in the estimated complete conditional, $\propdist$, can lead to inflated tails of the null $p$-values, causing a violation of the Type I or FDR error threshold. Here we develop a calibration technique that aims to produce conservative $p$-values by pessimistically weighting each null sample comparison.

We begin by casting \cref{alg:hrt} as an importance sampling scheme with $\propdist$ as a proposal distribution,
\begin{equation}
\label{eqn:approx_crt_p_value_2}
\begin{aligned}
\hat{p}_j &\approx& \frac{1}{1+\sum_{k=1}^K W^{(k)}} \left(1 + \sum_{k=1}^K \mathbbm{1}\left\{ \mathcal{G}(\mathcal{D}', \pi_{\hat{\theta}}) \geq \mathcal{G}(\widetilde{\mathcal{D}}^{(k)}, \pi_{\hat{\theta}}) \right\} W^{(k)} \right)& \, , \\
&& \widetilde{\mathcal{D}}^{(k)} = [\knockoff^{(k)}, \mathbf{X}_{\cdot -j}'] \, , \quad \knockoff^{(k)} \sim \propdist \, ,
\end{aligned}
\end{equation}
where $W^{(k)}$ is the importance weight,
\begin{equation}
\label{eqn:true_importance_weight}
W^{(k)} = \frac{\ccdist(\knockoff^{(k)})}{\propdist(\knockoff^{(k)})} \, ,
\end{equation}
where $\ccdist(\knockoff^{(k)}) = \prod_{i=1}^{n'} \ccdist(\widetilde{\mathbf{X}}_{ij}^{(k)})$, and analogously for $\propdist(\knockoff^{(k)})$.
The estimate in \eqref{eqn:approx_crt_p_value_2} is consistent, but it still relies on the unknown true conditional $\ccdist$, making it impossible to compute in practice.

Let $\lowerdist$ and $\upperdist$ be functions that bound the true probability of the null sample: $0 \leq \lowerdist(\knockoff) \leq \ccdist(\knockoff) \leq \upperdist(\knockoff)$.  If we have access to such a function, we can then bound $p_j$ in expectation by biasing the importance weights to be conservative. This works by choosing the importance weight dynamically based on the outcome of each test statistic comparison,
\begin{equation}
\label{eqn:crt_exact_bound}
\widehat{W}^{(k)} = \begin{cases}
\frac{\lowerdist(\knockoff^{(k)})}{\propdist(\knockoff^{(k)})} &\quad \text{if }\mathcal{G}(\mathcal{D}', \pi_{\hat{\theta}}) < \mathcal{G}(\widetilde{\mathcal{D}}^{(k)}, \pi_{\hat{\theta}}) \\
\frac{\upperdist(\knockoff^{(k)})}{\propdist(\knockoff^{(k)})} &\quad \text{otherwise.}
\end{cases}
\end{equation}
Under this conservative weighting, $\mathbb{E}\left[\widehat{p}_j\right] \geq p_j$. The tighter the bound from ($\lowerdist$, $\upperdist$), the less conservative the $p$-value estimate and thus the more powerful the test statistic. Trivially, for any true $\ccdist$, we can always choose $\lowerdist = 0$ and guarantee $\hat{p}$ will be conservative. It will always be $1$, however, so the method will be without power. To be effective, the bound must be relatively tight.



To generate approximate bounds on $\ccdist$, we treat the CDE model as a black box and use the bootstrap to estimate quantiles \citep[see ][for a discussion on the technical issues with using the bootstrap for confidence intervals]{efron:tibshirani:1998:problem-of-regions}. We create $b$ bootstrap resamples of the dataset and fit $b$ corresponding CDE models $\bootstrapdist = (\propdist^{(1)}, \propdist^{(2)}, \ldots, \propdist^{(b)})$, with $\propdist^{(1)}$ used as the proposal distribution. Given a choice of lower and upper quantile, $(l, u)$, we can replace the reweighting scheme in \eqref{eqn:crt_exact_bound} with an approximate one,
\begin{equation}
\label{eqn:crt_approx_bound}
\widehat{W}^{(k)} = \begin{cases}
\frac{\bootstrapdist^{(l)}(\knockoff^{(k)})}{\propdist^{(1)}(\knockoff^{(k)})} &\quad \text{if }\mathcal{G}(\mathcal{D}', \pi_{\hat{\theta}}) < \mathcal{G}(\widetilde{\mathcal{D}}^{(k)}, \pi_{\hat{\theta}}) \\
\frac{\bootstrapdist^{(u)}(\knockoff^{(k)})}{\propdist^{(1)}(\knockoff^{(k)})} &\quad \text{otherwise.}
\end{cases}
\end{equation}
where $\bootstrapdist^{(i)}$ denotes the $i^{\text{th}}$ quantile of the estimates in $\bootstrapdist$.

We find that taking the ratio of the products as in \cref{eqn:crt_approx_bound} is numerically unstable for large $n'$ and quickly goes to $0$ or $\infty$ (numerically). To stabilize the procedure, we take the geometric mean of the ratio of constituent-to-proposal conditional probability estimate $\mathbb{G}$ for each model across all $n'$ samples,
\begin{equation}
\label{eqn:geom_mean_weight}
\widehat{W}^{(k)} = \begin{cases}
\left(\displaystyle\prod_{i=1}^{n'}\frac{\bootstrapdist^{(l)}(\widetilde{X}_{ij}^{(k)})}{\propdist^{(1)}(\widetilde{X}_{ij}^{(k)})}\right)^{1/n'} &\quad \text{if }\mathcal{G}(\mathcal{D}', \pi_{\hat{\theta}}) < \mathcal{G}(\widetilde{\mathcal{D}}^{(k)}, \pi_{\hat{\theta}}) \\
\left(\displaystyle\prod_{i=1}^{n'}\frac{\bootstrapdist^{(u)}(\widetilde{X}_{ij}^{(k)})}{\propdist^{(1)}(\widetilde{X}_{ij}^{(k)})}\right)^{1/n'} &\quad \text{otherwise.}
\end{cases}
\end{equation}

Bootstrapping solves the first half of the calibration dilemma by providing a model-agnostic, nonparametric way to generate arbitrary quantile estimates. The other requirement is a way to choose the upper and lower quantiles. For this, we leverage the IID property of the conditional CDF values for each sample.

If a sampling distribution $\propdist$ is well-calibrated, the distribution of CDF values for all samples should be uniform. In practice, this does not happen when using highly-flexible black box conditional density estimators trained via maximum likelihood estimation. While many approaches to recalibration have been explored in the literature, most tend to apply to classification or regression models (e.g. Platt scaling \citep{platt:1999:calibration}). It is not immediately clear how one can calibrate quantiles for black box conditional density estimates. We use a data-adaptive heuristic described in \cref{app:one_way_ks} for continuous random variables; for other data types, we recommend simply using $(l, u) = (5, 95)$ as a conservative choice.


\section{Approximate calibration via a one-way KS statistic}
\label{app:one_way_ks}
For continuous random variables, we propose a one-way Kolmogorov-Smirnov (KS) statistic to choose the lower and upper quantiles of the bootstrapped models. An adaptation to discrete random variables is also available in our implementation. For other types of features, we recommend a conservative bound of $(l, u) = (5, 95)$; we find this is sufficient to control FDR in preliminary experiments.

We describe the approach in terms of finding a lower bound; the upper bound is determined analogously. To choose a lower quantile, we start with the median ($l=50$) and iteratively lower it until the CDF estimates dominate the uniform CDF from above in a Q-Q plot. Let $\bootstrapcdf$ be the CDF for $\bootstrapdist$. The one-way lower-KS statistic for $l$ is then,
\begin{equation}
\label{eqn:one_way_ks}
\text{KS}^{+}(l) = \underset{c \in [0,1]}{\text{maximum}}\left(0, c - \frac{1}{n} \sum_{i=1}^{n} \mathbb{I}\left[\bootstrapcdf^{(l)}(X_{ij}) \leq c \right]\right) \, .
\end{equation}
We then check if the current choice of $l$ yields a sufficiently-small test statistic. Specifically, we conduct a hypothesis test by Monte Carlo, where we repeatedly draw samples from $U(0,1)$ and only accept $l$ if it is smaller than $10^5$ draws from a true $U(0,1)$. This corresponds to solving the following optimization problem:
\begin{equation}
\begin{aligned}
\label{eqn:ks_optimization_criteria}
& \underset{l}{\text{argmax}} \quad & \text{max}(0, \text{KS}^{+}(l)) \\
& \text{subject to} & \quad \text{KS}^{+}(l) < \tau^{+} \, ,
\end{aligned}
\end{equation}
where $\tau^{+}$ is the MC-derived threshold; the upper quantile is found analgously. While the calibration procedure may seem very conservative, in practice it yields reasonable bounds even for the high-dimensional datasets in \cref{sec:benchmarks}.

\section{Conservativism of Bootstrap Heuristic}
With $\hat{W}^{(k)}$ as defined in \cref{eqn:crt_exact_bound},
relatively little is required for conservativism of the estimator of $p_j$ (the true $p$-value),
\begin{equation*}
\begin{aligned}
\hat{p}_j^K &\approx& \frac{1}{1+\sum_{k=1}^K \hat{W}^{(k)}} \left(1 + \sum_{k=1}^K \mathbbm{1}\left\{ \mathcal{G}(\mathcal{D}', \pi_{\hat{\theta}}) \geq \mathcal{G}(\widetilde{\mathcal{D}}^{(k)}, \pi_{\hat{\theta}}) \right\} \hat{W}^{(k)} \right)& \, , \\
&& \widetilde{\mathcal{D}}^{(k)} = [\knockoff^{(k)}, X_{\cdot -j}'] \, , \quad \knockoff^{(k)} \sim \propdist.
\end{aligned}
\end{equation*}
In particular, the functions $\lowerdist$ and $\upperdist$ only need to bound $\ccdist$ in a very crude sense.
Let $E_j = \left\{ \mathcal{G}(\mathcal{D}', \pi_{\hat{\theta}}) \geq \mathcal{G}(\widetilde{\mathcal{D}}^{(k)}, \pi_{\hat{\theta}}) \right\}$ be the event
that the empirical risk of a null sampler is less than the observed empirical risk. Then, 
\begin{theorem}
\label{thm:conservatism}
Let $\lowerdist$ and $\upperdist$ be such that $0 \le \EE_{\ccdist}\left[\frac{\lowerdist(\knockoff^{(k)})}{\ccdist(\knockoff^{(k)})}(1-\Ind \{E_j\})\right] \le 1-p_j$ and $p_j \le \EE_{\ccdist}\left[\frac{\upperdist(\knockoff^{(k)})}{\ccdist(\knockoff^{(k)})}\Ind \{E_j\}\right] < \infty$. Then, with importance weights as in \cref{eqn:crt_exact_bound}, $\lim_{K\to\infty}\hat{p}_j^K \ge p_j$ almost surely (with respect to the proposal distribution $Q_{j\given-j}$).
\end{theorem}

\begin{proof}
By the law of large numbers, almost surely,
\begin{align}
\lim_{K\to\infty}\hat{p}_j^K &= \frac{\EE_{Q_{j\given-j}}[\hat{W}^{(k)}\Ind \{E_j\}]}{\EE_{Q_{j\given-j}}[\hat{W}^{(k)}]}\\ 
  & = 1 / (1 + \frac{\EE_{Q_{j\given-j}}[\hat{W}^{(k)}(1-\Ind \{E_j\})]}{\EE_{Q_{j\given-j}}[\hat{W}^{(k)}\Ind \{E_j\}]}), \label{eq:suggestive_limit}
\end{align}
where the second line is an algebraic manipulation.
Notice that 
\begin{align*}
\EE_{Q_{j\given-j}}[\hat{W}^{(k)} \Ind\{E_j\}] &= \EE_{Q_{j\given-j}}[\frac{\upperdist(\knockoff^{(k)})}{\propdist(\knockoff^{(k)})} \Ind\{E_j\}]\\
& = \EE_{\ccdist}[\frac{\upperdist(\knockoff^{(k)})}{\ccdist(\knockoff^{(k)})} \Ind\{E_j\}] \\
& \ge p_j,
\end{align*}
where the last line is by assumption. Similarly, $\EE_{Q_{j\given-j}}[\hat{W}^{(k)}(1-\Ind \{E_j\})] \le 1-p_j$. 
The result then follows by comparing $p_j = 1/(1+\frac{1-p_j}{p_j})$ and \cref{eq:suggestive_limit}.
\end{proof}


\cref{thm:conservatism} shows that, because the test statistic is fixed, we require only that the upper and lower bounds are valid on average over the events $E_j$ and $1-E_j$. Note that our bootstrapping heuristic does not guarantee that \cref{thm:conservatism} will hold. Rather, it states that if the heuristic succeeds in obtaining bounds (on average) then the $p$-values will be conservative high probability.

\section{Additional benchmark results for known conditionals}
\label{app:benchmarks_known}
This section details additional results for the HRT algorithm and its extensions on the benchmarks in \cref{subsec:benchmarks:known}. First, we provide the corresponding false discovery rates for \cref{tab:benchmark_cvhrt,tab:benchmark_timing}. The results show that all experimental settings controlled the false discovery rate at or below the nominal $10\%$ level.

We next provide the results of using different predictive models: ordinary least squares (\cref{tab:app_benchmark_hrt_ols,tab:app_benchmark_fdr_ols}), lasso (\cref{tab:app_benchmark_hrt_lasso,tab:app_benchmark_fdr_lasso}), and Bayesian ridge regression (\cref{tab:app_benchmark_hrt_bayes,tab:app_benchmark_fdr_bayes}). These mirror the results using random forests presented in the main text.

\Cref{tab:app_benchmark_signals} provides the corresponding result for the lasso and OLS models with different signal strengths for the nonnull features.

\begin{figure}[th!]
\centering
\includegraphics[width=0.7\textwidth]{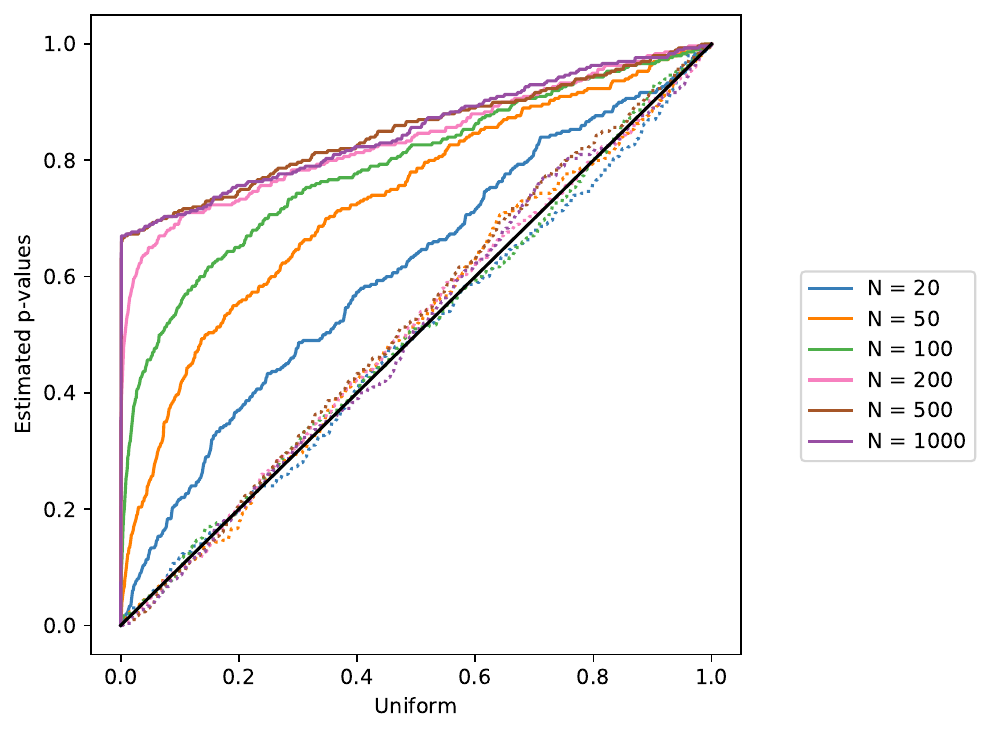}
\caption{\label{fig:qq_hrt} Q-Q plot for the HRT (Algorithm 1) simulations in \cref{subsec:benchmarks:known}. Solid lines show the CDFs of nonnull $p$-values; dotted lines show the CDFs of null $p$-values.}
\end{figure}

\begin{figure}[th!]
\centering
\includegraphics[width=0.7\textwidth]{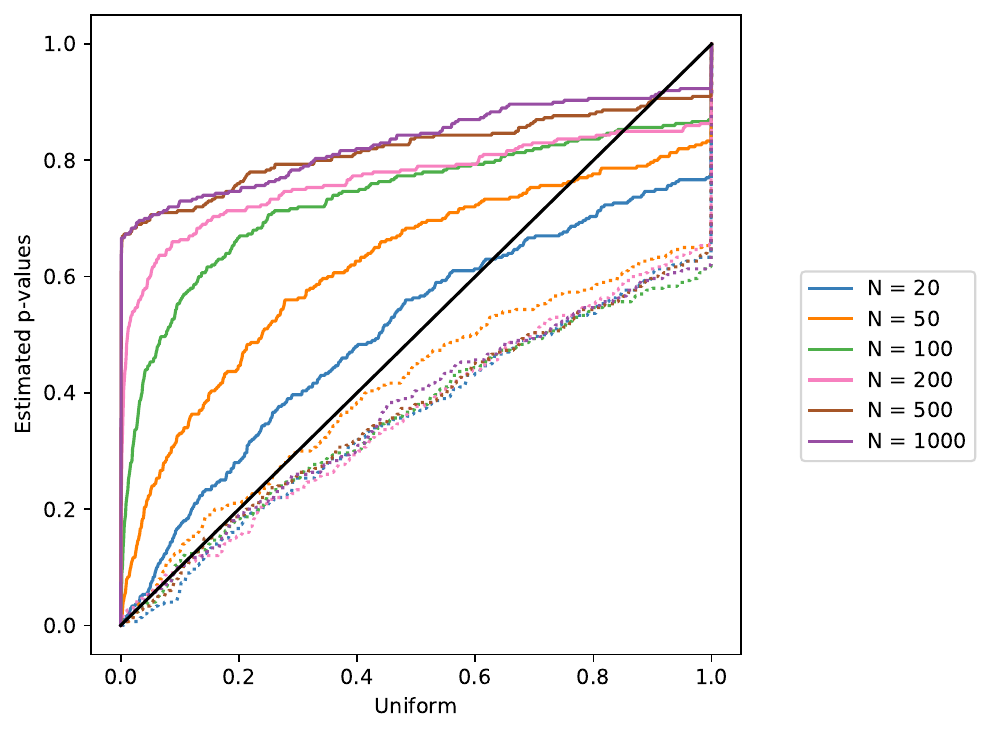}
\caption{\label{fig:qq_cvhrt} Q-Q plot for the CV-HRT (Algorithm 2) simulations in \cref{subsec:benchmarks:known}. Solid lines show the CDFs of nonnull $p$-values; dotted lines show the CDFs of null $p$-values.}
\end{figure}

\begin{figure}[th!]
\centering
\includegraphics[width=0.7\textwidth]{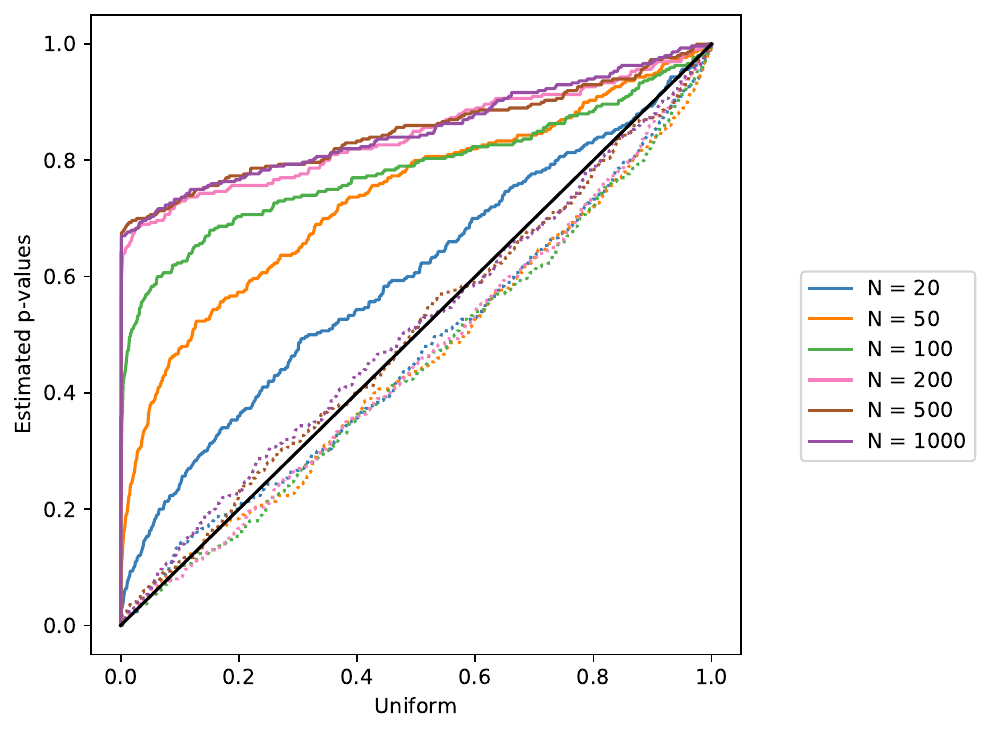}
\caption{\label{fig:qq_acvhrt} Q-Q plot for the approximate CV-HRT (Algorithm 3) simulations in \cref{subsec:benchmarks:known}. Solid lines show the CDFs of nonnull $p$-values; dotted lines show the CDFs of null $p$-values.}
\end{figure}

\begin{figure}[th!]
\centering
\includegraphics[width=0.7\textwidth]{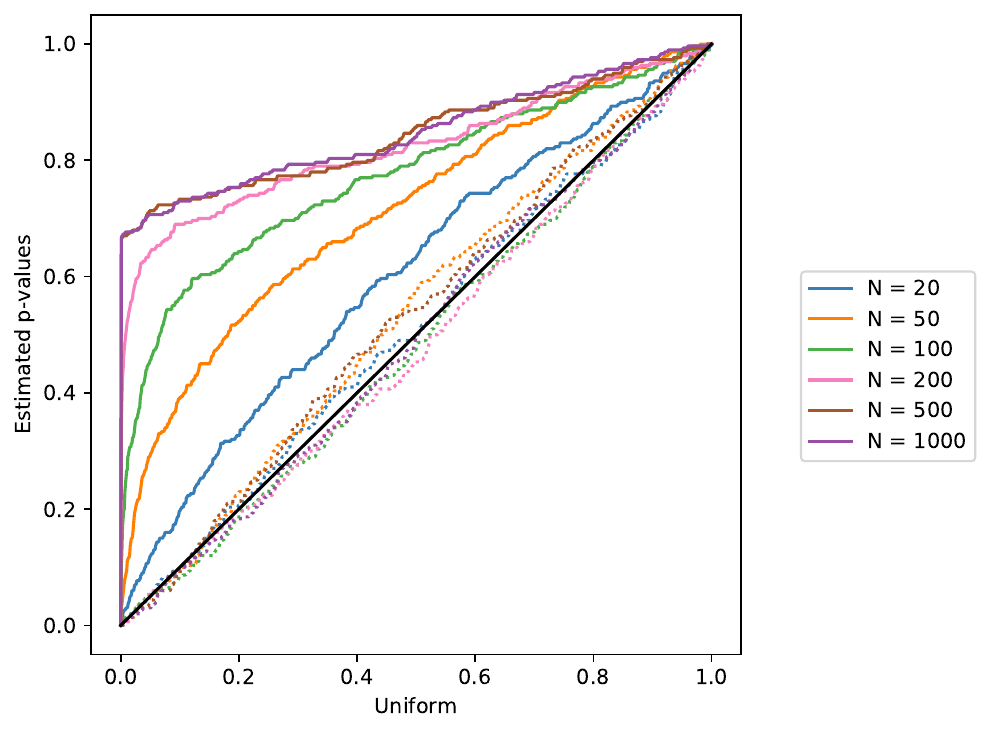}
\caption{\label{fig:qq_hpt} Q-Q plot for the HPT (Algorithm 4) simulations in \cref{subsec:benchmarks:known}. Solid lines show the CDFs of nonnull $p$-values; dotted lines show the CDFs of null $p$-values.}
\end{figure}

\begin{figure}[th!]
\centering
\includegraphics[width=0.7\textwidth]{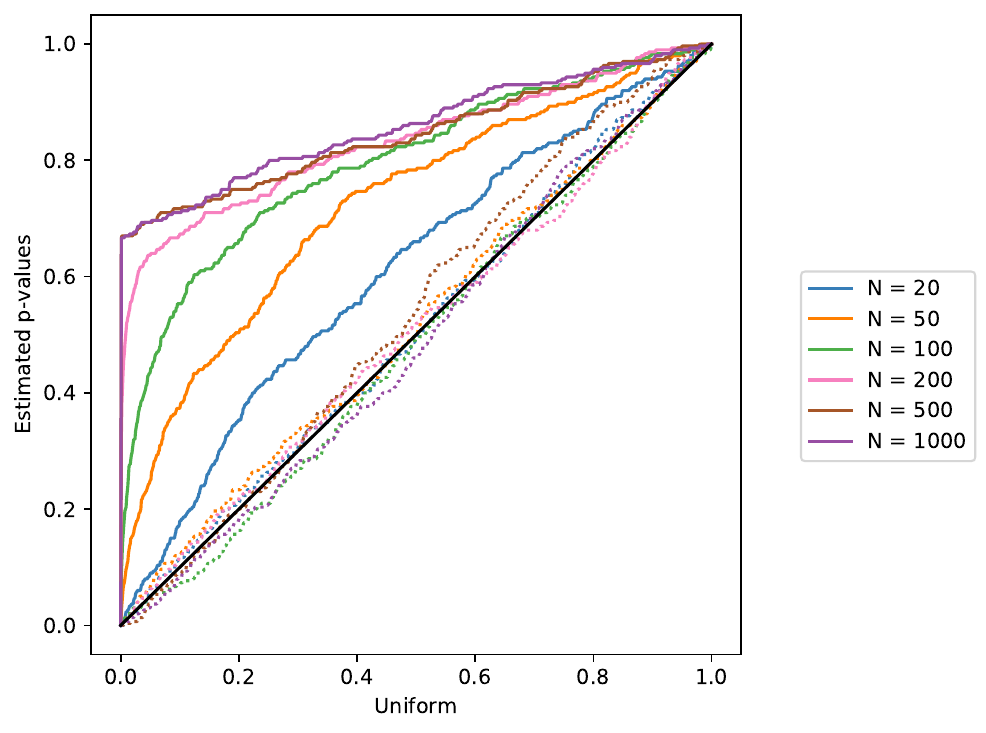}
\caption{\label{fig:qq_hgt} Q-Q plot for the HGT (Algorithm 5) simulations in \cref{subsec:benchmarks:known}. Solid lines show the CDFs of nonnull $p$-values; dotted lines show the CDFs of null $p$-values.}
\end{figure}

Finally, \cref{fig:qq_hrt,fig:qq_cvhrt,fig:qq_acvhrt,fig:qq_hpt,fig:qq_hgt} present the Q-Q plots for $p$-values generated by the HRT and the four proposed extension algorithms. For the two cross-validation algorithms, we select $M=5$ folds as an illustrative case. All results show that the methods generate empirically uniform or conservative null $p$-values (dotted lines in the figures). Further, the nonnull $p$-value distributions (solid lines) shift progressively to the left as the sample size grows.

\begin{table}[t!]
\centering
\begin{tabular}{lccccccc}
\toprule
Model & Splits                    & $n=20$    & $n=50$    & $n=100$  & $n=200$   & $n=500$   & $n=1000$  \\
\midrule
HRT (\cref{alg:hrt}) & 1        & 0.05 & 0.03 & 0.05  & 0.04  & 0.06  & 0.06    \\
\midrule
\multirow{6}{*}{\makecell{CV-HRT \\ (\cref{alg:cvhrt-valid})}} 
& 2  &  0.03 & 0.05 & 0.03  & 0.06  & 0.07  & 0.05   \\
& 3  &  0.04 & 0.03 & 0.06  & 0.05  & 0.05  & 0.05   \\
& 4  &  0.08 & 0.06 & 0.03  & 0.05  & 0.03  & 0.03   \\
& 5  &  0.01 & 0.05 & 0.06  & 0.06  & 0.03  & 0.05   \\
& 10 &  0.02 & 0.06 & 0.04  & 0.07  & 0.05  & 0.06   \\
& 20 &  0    & 0    & 0.04  & 0.06  & 0.05  & 0.07  \\
\midrule
\multirow{6}{*}{\makecell{CV-HRT \\ (\cref{alg:cvhrt-approx})}} 
& 2  & 0.05 & 0.06 & 0.06  & 0.06  & 0.06  & 0.06   \\
& 3  & 0.04 & 0.06 & 0.05  & 0.08  & 0.07  & 0.06   \\
& 4  & 0.07 & 0.06 & 0.05  & 0.04  & 0.04  & 0.07   \\
& 5  & 0.07 & 0.06 & 0.05  & 0.05  & 0.06  & 0.07   \\
& 10 & 0.07 & 0.03 & 0.09  & 0.05  & 0.04  & 0.03   \\
& 20 & 0.05 & 0.04 & 0.05  & 0.04  & 0.06  & 0.06   \\ 
\bottomrule
\end{tabular}
\caption{\label{tab:app_benchmark_fdr_rf} False discovery rate for the random forest predictive model in \cref{subsec:benchmarks:known}.}
\end{table}

\begin{table}[t!]
\centering
\begin{tabular}{cccccccc}
\toprule
& Model      & $n=20$    & $n=50$    & $n=100$  & $n=200$   & $n=500$   & $n=1000$  \\
\midrule
\multirow{3}{*}{FDR} &  HRT (\cref{alg:hrt})  & 0.07 & 0.04 & 0.05 & 0.04 & 0.03 & 0.04    \\
 &  HPT (\cref{alg:hpt})  & 0.06 & 0.07 & 0.05 & 0.07 & 0.04 & 0.03   \\
 &  HGT (\cref{alg:hgt}) & 0.06 & 0.03 & 0.05 & 0.05 & 0.04 & 0.03   \\
\bottomrule
\end{tabular}
\caption{\label{tab:app_benchmark_fdr} False discovery rate for the HRT, HPT, and HGT using random forests as the predictive model.}
\end{table}

\begin{table}[ht]
\centering
\begin{tabular}{lccccccc}
\toprule
Model & Splits                    & $n=20$    & $n=50$    & $n=100$  & $n=200$   & $n=500$   & $n=1000$  \\
\midrule
HRT (\cref{alg:hrt}) & 1        & 0.03 & 0.31 & 0.59  & 0.7   & 0.71  & 0.75    \\
\midrule
\multirow{6}{*}{\makecell{CV-HRT \\ (\cref{alg:cvhrt-valid})}} 
& 2  &  0.04 & 0.36 & 0.67  & 0.69  & 0.74  & 0.77   \\
& 3  &  0.08 & 0.3  & 0.64  & 0.7   & 0.69  & 0.75   \\
& 4  &  0.06 & 0.33 & 0.63  & 0.7   & 0.72  & 0.72   \\
& 5  &  0.05 & 0.28 & 0.51  & 0.68  & 0.73  & 0.73   \\
& 10 &  0.04 & 0.15 & 0.35  & 0.56  & 0.71  & 0.71   \\
& 20 &  0.02 & 0.08 & 0.15  & 0.42  & 0.68  & 0.71   \\
\midrule
\multirow{6}{*}{\makecell{CV-HRT \\ (\cref{alg:cvhrt-approx})}} 
& 2  & 0.04 & 0.43 & 0.68  & 0.72  & 0.74  & 0.75   \\
& 3  & 0.09 & 0.48 & 0.67  & 0.69  & 0.71  & 0.75   \\
& 4  & 0.15 & 0.53 & 0.68  & 0.7   & 0.72  & 0.76   \\
& 5  & 0.17 & 0.54 & 0.68  & 0.69  & 0.74  & 0.79   \\
& 10 & 0.14 & 0.52 & 0.66  & 0.69  & 0.72  & 0.78   \\
& 20 & 0.12 & 0.54 & 0.67  & 0.68  & 0.73  & 0.78   \\ 
\bottomrule
\end{tabular}
\caption{\label{tab:app_benchmark_hrt_ols} True positive rate for the ordinary least squares predictive model in \cref{subsec:benchmarks:known}.}
\end{table}

\begin{table}[ht]
\centering
\begin{tabular}{lccccccc}
\toprule
Model & Splits                    & $n=20$    & $n=50$    & $n=100$  & $n=200$   & $n=500$   & $n=1000$  \\
\midrule
HRT (\cref{alg:hrt}) & 1        & 0.08 & 0.28 & 0.61  & 0.69  & 0.7   & 0.72    \\
\midrule
\multirow{6}{*}{\makecell{CV-HRT \\ (\cref{alg:cvhrt-valid})}} 
& 2  &  0.06 & 0.35 & 0.65  & 0.69  & 0.71  & 0.73   \\
& 3  &  0.09 & 0.36 & 0.62  & 0.68  & 0.72  & 0.72   \\
& 4  &  0.04 & 0.24 & 0.58  & 0.66  & 0.69  & 0.69   \\
& 5  &  0.06 & 0.24 & 0.51  & 0.67  & 0.69  & 0.73   \\
& 10 &  0.01 & 0.15 & 0.32  & 0.62  & 0.68  & 0.71   \\
& 20 &  0.04 & 0.04 & 0.17  & 0.36  & 0.67  & 0.69   \\
\midrule
\multirow{6}{*}{\makecell{CV-HRT \\ (\cref{alg:cvhrt-approx})}} 
& 2  & 0.1  & 0.42 & 0.67  & 0.7   & 0.72  & 0.73   \\
& 3  & 0.12 & 0.48 & 0.66  & 0.7   & 0.71  & 0.73   \\
& 4  & 0.09 & 0.46 & 0.65  & 0.7   & 0.7   & 0.71   \\
& 5  & 0.13 & 0.47 & 0.66  & 0.69  & 0.71  & 0.72   \\
& 10 & 0.17 & 0.49 & 0.67  & 0.68  & 0.71  & 0.73   \\
& 20 & 0.19 & 0.51 & 0.68  & 0.69  & 0.71  & 0.75   \\ 
\bottomrule
\end{tabular}
\caption{\label{tab:app_benchmark_hrt_lasso} True positive rate for the lasso predictive model in \cref{subsec:benchmarks:known}.}
\end{table}

\begin{table}[t!]
\centering
\begin{tabular}{lccccccc}
\toprule
Model & Splits                    & $n=20$    & $n=50$    & $n=100$  & $n=200$   & $n=500$   & $n=1000$  \\
\midrule
HRT (\cref{alg:hrt}) & 1        & 0.08 & 0.35 & 0.6   & 0.7   & 0.69  & 0.72    \\
\midrule
\multirow{6}{*}{\makecell{CV-HRT \\ (\cref{alg:cvhrt-valid})}} 
& 2  &  0.09 & 0.43 & 0.68  & 0.69  & 0.7   & 0.72   \\
& 3  &  0.1  & 0.34 & 0.63  & 0.69  & 0.68  & 0.73   \\
& 4  &  0.08 & 0.35 & 0.62  & 0.68  & 0.72  & 0.7    \\
& 5  &  0.08 & 0.31 & 0.53  & 0.67  & 0.71  & 0.72   \\
& 10 &  0.04 & 0.16 & 0.33  & 0.57  & 0.69  & 0.7    \\
& 20 &  0.02 & 0.06 & 0.15  & 0.44  & 0.68  & 0.7   \\
\midrule
\multirow{6}{*}{\makecell{CV-HRT \\ (\cref{alg:cvhrt-approx})}} 
& 2  & 0.12 & 0.47 & 0.67  & 0.7   & 0.7   & 0.73   \\
& 3  & 0.13 & 0.5  & 0.67  & 0.68  & 0.7   & 0.73   \\
& 4  & 0.18 & 0.52 & 0.68  & 0.69  & 0.7   & 0.74   \\
& 5  & 0.14 & 0.55 & 0.67  & 0.69  & 0.73  & 0.76   \\
& 10 & 0.15 & 0.55 & 0.67  & 0.68  & 0.71  & 0.76   \\
& 20 & 0.17 & 0.53 & 0.68  & 0.68  & 0.72  & 0.76   \\ 
\bottomrule
\end{tabular}
\caption{\label{tab:app_benchmark_hrt_bayes} True positive rate for the Bayesian ridge predictive model in \cref{subsec:benchmarks:known}.}
\end{table}

\begin{table}[t!]
\centering
\begin{tabular}{lccccccc}
\toprule
Model & Splits                    & $n=20$    & $n=50$    & $n=100$  & $n=200$   & $n=500$   & $n=1000$  \\
\midrule
HRT (\cref{alg:hrt}) & 1        & 0.06 & 0.05 & 0.05  & 0.06  & 0.06  & 0.06    \\
\midrule
\multirow{6}{*}{\makecell{CV-HRT \\ (\cref{alg:cvhrt-valid})}} 
& 2  &  0.05 & 0.06 & 0.05  & 0.06  & 0.03  & 0.06   \\
& 3  &  0.04 & 0.05 & 0.02  & 0.06  & 0.04  & 0.06   \\
& 4  &  0.06 & 0.05 & 0.05  & 0.04  & 0.05  & 0.04   \\
& 5  &  0.05 & 0.02 & 0.05  & 0.04  & 0.06  & 0.04   \\
& 10 &  0.04 & 0.05 & 0.08  & 0.06  & 0.05  & 0.07   \\
& 20 &  0.04 & 0.04 & 0.03  & 0.06  & 0.05  & 0.04   \\
\midrule
\multirow{6}{*}{\makecell{CV-HRT \\ (\cref{alg:cvhrt-approx})}} 
& 2  & 0.05 & 0.05 & 0.04  & 0.05  & 0.06  & 0.05   \\
& 3  & 0.06 & 0.05 & 0.04  & 0.05  & 0.04  & 0.05   \\
& 4  & 0.08 & 0.06 & 0.05  & 0.05  & 0.04  & 0.03   \\
& 5  & 0.07 & 0.03 & 0.03  & 0.03  & 0.04  & 0.04   \\
& 10 & 0.07 & 0.03 & 0.05  & 0.04  & 0.05  & 0.04   \\
& 20 & 0.05 & 0.05 & 0.04  & 0.04  & 0.03  & 0.04   \\ 
\bottomrule
\end{tabular}
\caption{\label{tab:app_benchmark_fdr_ols} False discovery rate for the ordinary least squares predictive model in \cref{subsec:benchmarks:known}.}
\end{table}

\begin{table}[t!]
\centering
\begin{tabular}{lccccccc}
\toprule
Model & Splits                    & $n=20$    & $n=50$    & $n=100$  & $n=200$   & $n=500$   & $n=1000$  \\
\midrule
HRT (\cref{alg:hrt}) & 1        & 0    & 0.01 & 0.02  & 0.02  & 0.01  & 0.02    \\
\midrule
\multirow{6}{*}{\makecell{CV-HRT \\ (\cref{alg:cvhrt-valid})}} 
& 2  &  0    & 0.02 & 0.02  & 0.03  & 0.01  & 0.02   \\
& 3  &  0.02 & 0    & 0.02  & 0.01  & 0.01  & 0.03   \\
& 4  &  0.02 & 0    & 0.01  & 0.01  & 0.01  & 0.01   \\
& 5  &  0.03 & 0.02 & 0.02  & 0.01  & 0.02  & 0.02   \\
& 10 &  0.02 & 0.02 & 0.02  & 0.02  & 0.01  & 0.04   \\
& 20 &  0    & 0.02 & 0.02  & 0.02  & 0.02  & 0.03   \\
\midrule
\multirow{6}{*}{\makecell{CV-HRT \\ (\cref{alg:cvhrt-approx})}} 
& 2  & 0.02 & 0.02 & 0.03  & 0.03  & 0.01  & 0.02   \\
& 3  & 0.01 & 0.03 & 0.02  & 0.02  & 0.02  & 0.04   \\
& 4  & 0.04 & 0.03 & 0.03  & 0.01  & 0.01  & 0.02   \\
& 5  & 0.02 & 0.02 & 0.01  & 0.03  & 0.04  & 0.03   \\
& 10 & 0.03 & 0.06 & 0.02  & 0.02  & 0.02  & 0.04   \\
& 20 & 0.02 & 0.03 & 0.02  & 0.02  & 0.03  & 0.02   \\ 
\bottomrule
\end{tabular}
\caption{\label{tab:app_benchmark_fdr_lasso} False discovery rate for the lasso predictive model in \cref{subsec:benchmarks:known}.}
\end{table}

\begin{table}[t!]
\centering
\begin{tabular}{lccccccc}
\toprule
Model & Splits                    & $n=20$    & $n=50$    & $n=100$  & $n=200$   & $n=500$   & $n=1000$  \\
\midrule
HRT (\cref{alg:hrt}) & 1        & 0.06 & 0.04 & 0.04  & 0.03  & 0.03  & 0.02    \\
\midrule
\multirow{6}{*}{\makecell{CV-HRT \\ (\cref{alg:cvhrt-valid})}} 
& 2  &  0.04 & 0.02 & 0.05  & 0.04  & 0.01  & 0.02   \\
& 3  &  0.06 & 0.05 & 0.01  & 0.04  & 0.01  & 0.02   \\
& 4  &  0.04 & 0.04 & 0.04  & 0.03  & 0.02  & 0.01   \\
& 5  &  0.01 & 0.03 & 0.04  & 0.03  & 0.02  & 0.02   \\
& 10 &  0.09 & 0.03 & 0.03  & 0.04  & 0.03  & 0.02   \\
& 20 &  0.03 & 0.02 & 0.02  & 0.04  & 0.02  & 0.01  \\
\midrule
\multirow{6}{*}{\makecell{CV-HRT \\ (\cref{alg:cvhrt-approx})}} 
& 2  & 0.05 & 0.03 & 0.02  & 0.04  & 0.03  & 0.03   \\
& 3  & 0.04 & 0.02 & 0.02  & 0.03  & 0.03  & 0.03   \\
& 4  & 0.06 & 0.03 & 0.03  & 0.03  & 0.02  & 0.02   \\
& 5  & 0.06 & 0.01 & 0.02  & 0.01  & 0.03  & 0.02   \\
& 10 & 0.02 & 0.01 & 0.03  & 0.01  & 0.02  & 0.02   \\
& 20 & 0.05 & 0.02 & 0.02  & 0.02  & 0.02  & 0.02   \\ 
\bottomrule
\end{tabular}
\caption{\label{tab:app_benchmark_fdr_bayes} False discovery rate for the Bayesian ridge predictive model in \cref{subsec:benchmarks:known}.}
\end{table}

\begin{table}[t!]
\centering
\begin{tabular}{cccccc}
\toprule
\multirow{2}{*}{Signal Strength} & \multirow{2}{*}{Algorithm} & \multicolumn{2}{c}{TPR} & \multicolumn{2}{c}{FDR} \\
& & OLS & Lasso & OLS & Lasso \\
\midrule
\multirow{3}{*}{$0.1$} & HRT (\cref{alg:hrt}) & 0.55 & 0.58 & 0.03 & 0.01 \\
& CV-HRT (\cref{alg:cvhrt-valid}) & 0.69 & 0.66 & 0.05 & 0.03 \\
& CV-HRT (\cref{alg:cvhrt-approx}) & 0.69 & 0.68 & 0.04 & 0.04 \\
\midrule
\multirow{3}{*}{$0.2$} & HRT (\cref{alg:hrt}) & 0.70 & 0.69 & 0.02 & 0.01 \\
& CV-HRT (\cref{alg:cvhrt-valid}) & 0.73 & 0.69 & 0.03 & 0.02 \\
& CV-HRT (\cref{alg:cvhrt-approx}) & 0.72 & 0.73 & 0.05 & 0.02 \\
\midrule
\multirow{3}{*}{$0.3$} & HRT (\cref{alg:hrt}) & 0.71 & 0.71 & 0.03 & 0.01 \\
& CV-HRT (\cref{alg:cvhrt-valid}) & 0.76 & 0.72 & 0.04 & 0.02 \\
& CV-HRT (\cref{alg:cvhrt-approx}) & 0.76 & 0.76 & 0.05 & 0.02 \\
\midrule
\multirow{3}{*}{$0.4$} & HRT (\cref{alg:hrt}) & 0.71 & 0.72 & 0.03 & 0.01 \\
& CV-HRT (\cref{alg:cvhrt-valid}) & 0.75 & 0.72 & 0.04 & 0.02 \\
& CV-HRT (\cref{alg:cvhrt-approx}) & 0.80 & 0.80 & 0.06 & 0.02 \\
\midrule
\multirow{3}{*}{$0.5$} & HRT (\cref{alg:hrt}) & 0.72 & 0.72 & 0.04 & 0.01 \\
& CV-HRT (\cref{alg:cvhrt-valid}) & 0.76 & 0.75 & 0.02 & 0.02 \\
& CV-HRT (\cref{alg:cvhrt-approx}) & 0.83 & 0.82 & 0.05 & 0.02 \\
\midrule
\multirow{3}{*}{$0.6$} & HRT (\cref{alg:hrt}) & 0.75 & 0.72 & 0.05 & 0.02 \\
& CV-HRT (\cref{alg:cvhrt-valid}) & 0.77 & 0.77 & 0.03 & 0.01 \\
& CV-HRT (\cref{alg:cvhrt-approx}) & 0.85 & 0.85 & 0.05 & 0.02 \\
\midrule
\multirow{3}{*}{$0.7$} & HRT (\cref{alg:hrt}) & 0.74 & 0.75 & 0.06 & 0.03 \\
& CV-HRT (\cref{alg:cvhrt-valid}) & 0.77 & 0.78 & 0.03 & 0.02 \\
& CV-HRT (\cref{alg:cvhrt-approx}) & 0.87 & 0.88 & 0.04 & 0.03 \\
\midrule
\multirow{3}{*}{$0.8$} & HRT (\cref{alg:hrt}) & 0.76 & 0.75 & 0.06 & 0.03 \\
& CV-HRT (\cref{alg:cvhrt-valid}) & 0.79 & 0.79 & 0.03 & 0.02 \\
& CV-HRT (\cref{alg:cvhrt-approx}) & 0.90 & 0.89 & 0.03 & 0.04 \\
\midrule
\multirow{3}{*}{$0.9$} & HRT (\cref{alg:hrt}) & 0.77 & 0.77 & 0.06 & 0.03 \\
& CV-HRT (\cref{alg:cvhrt-valid}) & 0.83 & 0.80 & 0.03 & 0.02 \\
& CV-HRT (\cref{alg:cvhrt-approx}) & 0.93 & 0.90 & 0.03 & 0.03 \\
\midrule
\multirow{3}{*}{$0.9$} & HRT (\cref{alg:hrt}) & 0.78 & 0.78 & 0.06 & 0.03 \\
& CV-HRT (\cref{alg:cvhrt-valid}) & 0.83 & 0.82 & 0.03 & 0.02 \\
& CV-HRT (\cref{alg:cvhrt-approx}) & 0.94 & 0.92 & 0.03 & 0.03 \\
\bottomrule
\end{tabular}
\caption{\label{tab:app_benchmark_signals} OLS and lasso results as the signal strength of the features changes. All simulations used $n=200$ data points and $M=5$ folds for the cross-validation methods.}
\end{table}

\section{Additional benchmark results for unknown conditionals}
\label{app:benchmarks_unknown}

\begin{figure}[t]
\centering
\begin{subfigure}{0.48\textwidth}\hspace{0.065\textwidth}\includegraphics[width=0.97\textwidth]{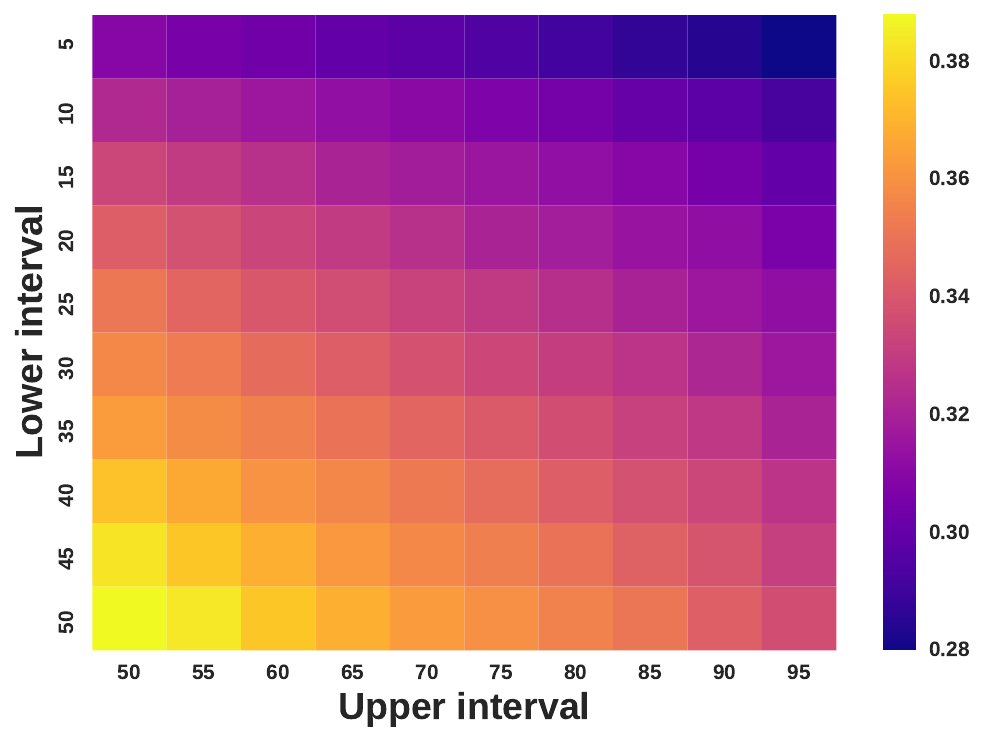}\caption{TPR}\end{subfigure}
\begin{subfigure}{0.48\textwidth}\hspace{0.065\textwidth}\includegraphics[width=0.97\textwidth]{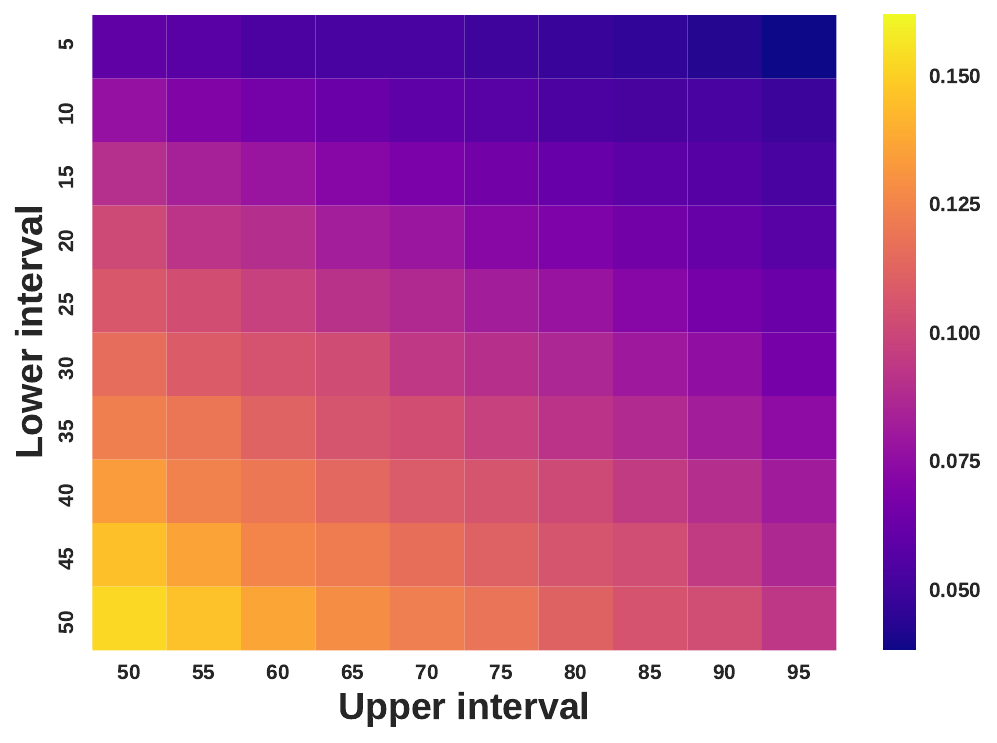}\caption{Average FDP}\end{subfigure}
\begin{subfigure}{0.45\textwidth}\includegraphics[width=\textwidth]{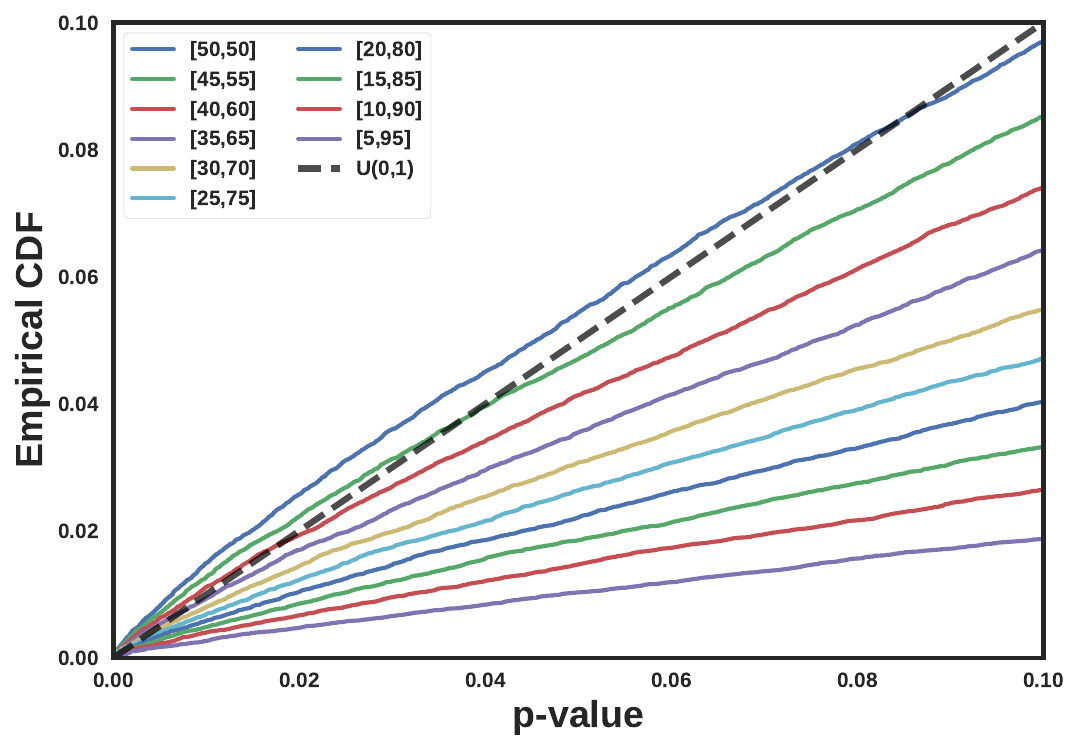}\caption{Null p-values}\end{subfigure}
\begin{subfigure}{0.45\textwidth}\includegraphics[width=\textwidth]{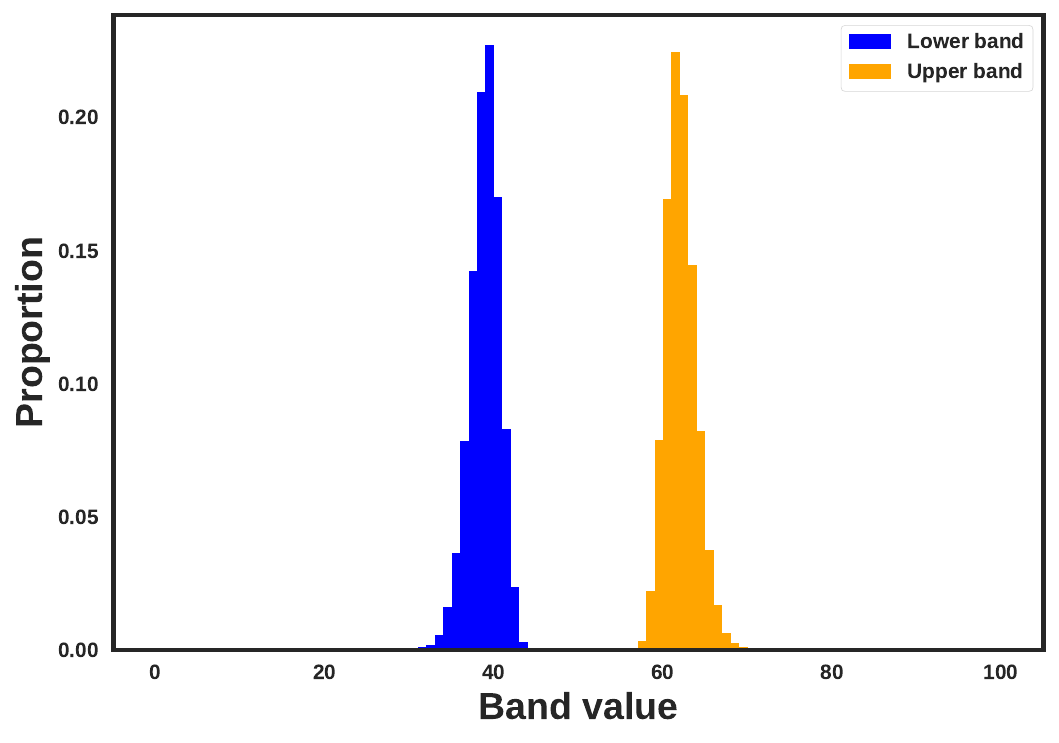}\caption{Calibration bounds}\end{subfigure}
\caption{\label{fig:calibration} (a) TPR and (b) empirical FDR for different fixed upper and lower quantile intervals on the benchmarks. (c) CDF of the true null $p$-values across all benchmark trials for different interval selections. (d) distribution of bounds selected by the calibration technique.}
\end{figure}

\cref{fig:calibration} provides a closer look at the calibration procedure results for the cross-validation HRT. The top two panels show a sweep of possible choices of the upper and lower bounds for both power and empirical FDR. Even at a stringent $90\%$ confidence interval (top right corner of both heat maps), power is still $28\%$; empirical FDR declines relatively rapidly, being only $\approx5\%$ at a $90\%$ interval. The bottom left panel shows the the CDF values of the $p$-values for the null features using different bounds. The panel is zoomed in to show the $[0,0.1]$ region where $p$-values may be rejected under the target FDR threshold. Without any calibration, even the bootstrapped median (i.e. the $[50, 50]$ interval) still shows an inflation in the lower portion of the CDF, leading to invalid null $p$-values. A $20$--$30\%$ confidence interval biases the null $p$-values down sufficiently to control the empirical FDR near the target level. The bottom right panel shows the distribution of bounds chosen by the data-adaptive procedure in each of the $100$ independent trials; the procedure chooses roughly a $25\%$ confidence region for most trials. 

\section{Benchmark results with BY correction}
\label{app:by_benchmarks}
\begin{figure}[t]
\centering
\begin{subfigure}{0.95\textwidth}\includegraphics[width=\textwidth]{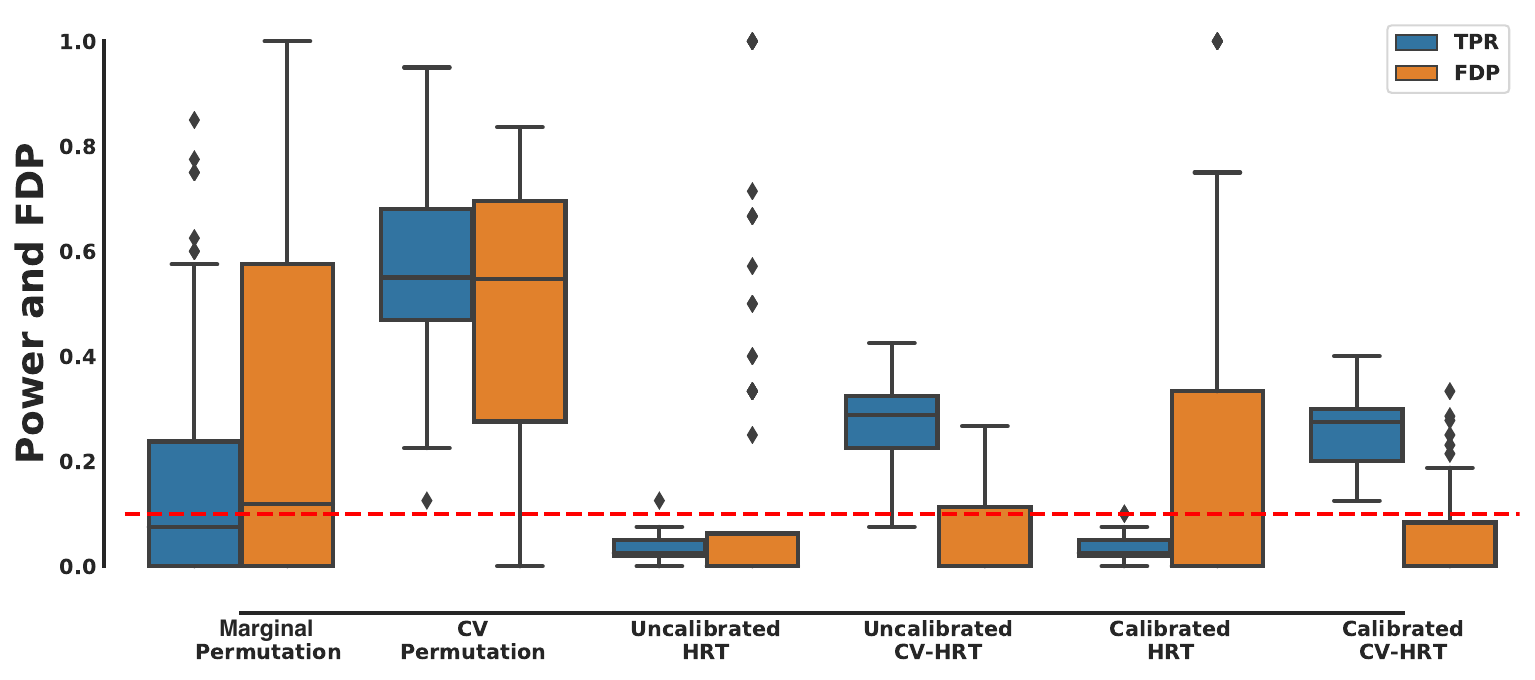}\end{subfigure}
\begin{subfigure}{0.95\textwidth}\includegraphics[width=\textwidth]{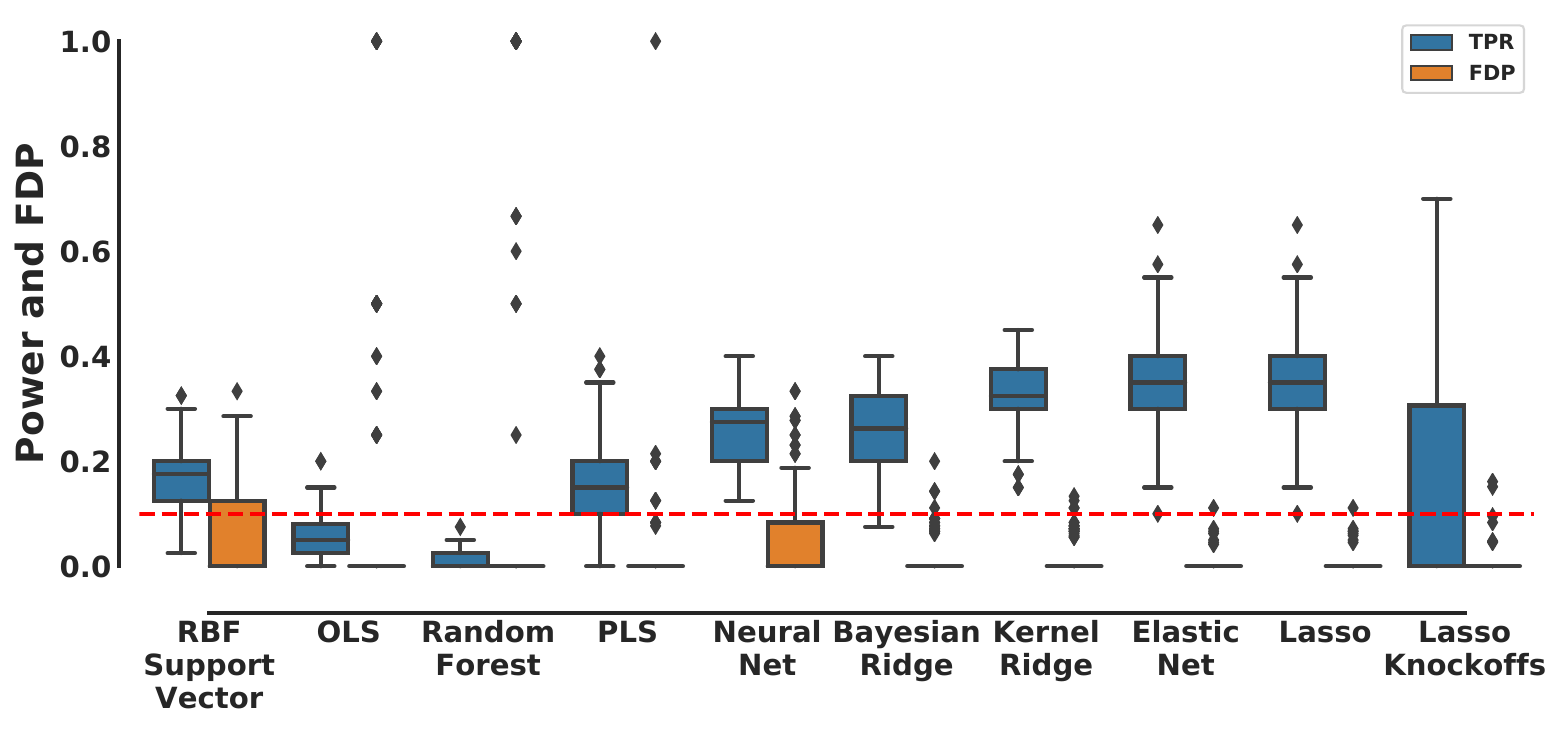}\end{subfigure}
\caption{\label{fig:by_benchmark_results} Top: Power and FDP results for each HRT variant on the benchmark simulation. Bottom: Power and FDP for different choices of predictive model used in the CV-HRT, ordered by empirical risk of the predictive model. The right-most result is for a lasso model using model-X knockoffs for feature selection. All results use BY correction rather than simple BH.}
\end{figure}

\cref{fig:by_benchmark_results} shows the results of the same benchmarks from \cref{sec:benchmarks}, but using Benjamini-Yekutieli (BY) correction \citep{benjamini:yekutieli:2001:dependence} instead of BH correction. Although the BY correction is technically necessary to ensure FDR control, the results show that for our simulation the actual correction is conservative. Nevertheless, the BY-corrected HRT p-values still have high power in our simulations.

\section{Disentangling knockoffs from empirical risk testing}
\label{app:eross}
In our experiments, the HRT has higher power than the lasso knockoffs of \citet{candes:etal:2018:panning}. Intuitively, there could be two reasons for this power. First, the HRT computes $p$-values and performs selection using BH; knockoffs work in a single-shot setting with simultaneous selection via a martingale-based filter \citep{barber:candes:2015:knockoffs}. Second, the lasso knockoffs use a coefficient magnitude test statistic rather than held out empirical risk. Either of these differences in procedures could conceivably be causing the discrepancy in power.

To investigate this discrepancy, we consider a hybrid knockoff-HRT heuristic approach. The hybrid approach uses the cross-validation empirical risk test statistic of the CV-HRT, but performs selection via the knockoff filter. \cref{alg:erk} presents the full algorithm.

\begin{algorithm}[t]
\caption{\label{alg:erk} A knockoff-style empirical risk selection heuristic}
\begin{algorithmic}[1]
\Procedure{Hybrid}{training data $\mathcal{D}$, test data $\mathcal{D}'$, model $\pi$, training objective $\mathcal{L}_{\pi}$, empirical risk function $\mathcal{G}$, FDR threshold $\alpha$}
\State Fit $\hat{\theta}$ by optimizing $\mathcal{L}_{\pi}(\mathcal{D}, \theta)$.
\State Compute the empirical risk on held out data, $t \leftarrow \mathcal{G}(\mathcal{D}', \pi_{\hat{\theta}})$.
\For {each feature $j$}
    \State Sample $\widetilde{X}_{\cdot j}' \sim \ccdist$.
    \State Create a new dataset $\widetilde{\mathcal{D}}$ by replacing $X_{\cdot j}'$ in $\mathcal{D}'$ with $\knockoff'$.
    \State Compute the change in empirical risk, $w^{(j)} \leftarrow \mathcal{G}(\widetilde{\mathcal{D}}, \pi_{\hat{\theta}})$ - t.
\EndFor
\State $\omega^* \leftarrow \underset{\omega \geq 0}{\text{min.}}\text{ } \omega \, , \text{subject to } \left[ \frac{1+\text{\# } w^{(j)} \leq -\omega}{\text{\# } w^{(j)} \geq \omega} \leq \alpha \right] \, .$
\Return{ discoveries at the $\alpha$ level: $\{ j \colon w^{(j)} \geq \omega^*\}$.}
\EndProcedure
\end{algorithmic}
\end{algorithm}

Unlike the HRT and knockoffs, this hybrid approach does not produce valid test statistics, even with the true complete conditionals. The knockoff filter requires that the signs of each test statistic are i.i.d. coin flips under the null. The hybrid approach, by only conditioning on the complete conditionals and not on the joint, creates a dependency between all of the signs. Thus, applying the knockoff filter to the one-shot empirical risk statistics is not guaranteed to control FDR. The closely related ``swap'' knockoff statistic \citep{gimenez:etal:2018:gmm-knockoffs} would be a valid knockoff approach that uses empirical risk, but it would require changing the predictive model. Since we are only interested in determining power here, FDR control is less important than retaining the same predictive model.

\begin{figure}[t]
\centering
\includegraphics[width=\textwidth]{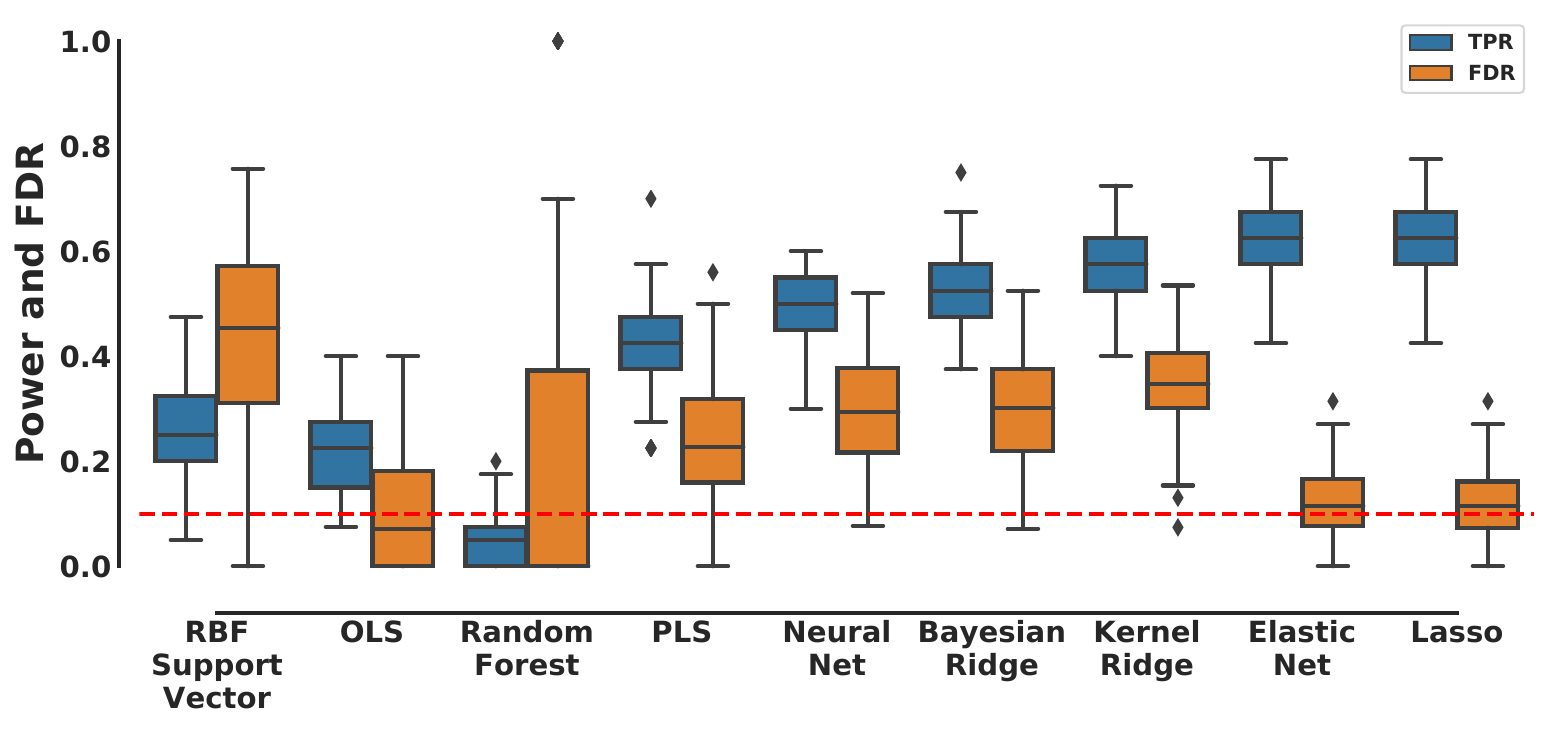}
\caption{\label{fig:knockoff_benchmark_results} Power and FDP for different choices of predictive model used with \cref{alg:erk}, ordered by empirical risk of the predictive model.}
\end{figure}

\begin{figure}[t]
\centering
\includegraphics[width=0.7\textwidth]{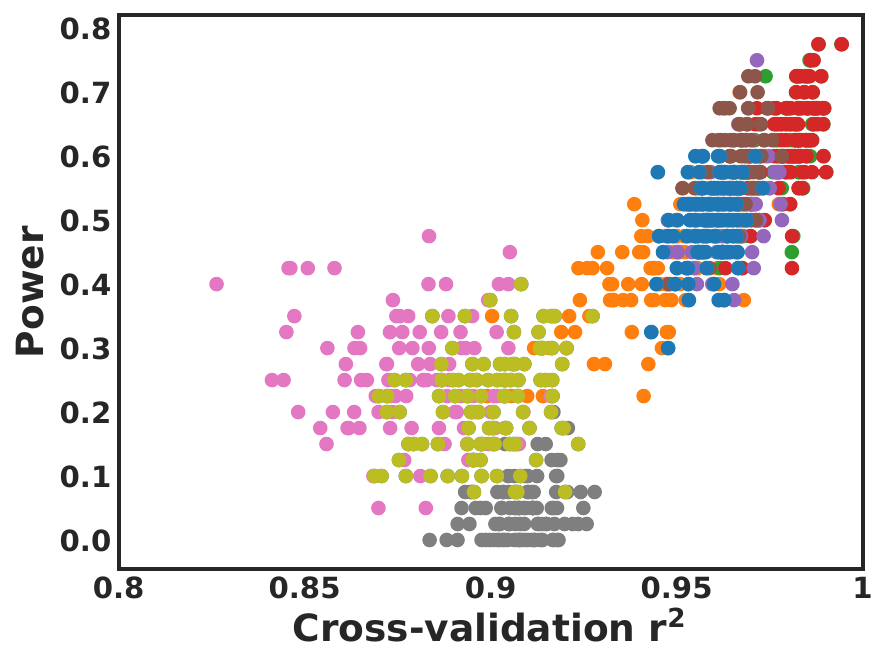}
\caption{\label{fig:knockoffs_benchmarks_r2} Predictive model performance versus feature selection power for each model and independent trial using \cref{alg:erk}.}
\end{figure}

\cref{fig:knockoff_benchmark_results} shows the results of the same simulations, but using \cref{alg:erk} instead of the HRT. The method maintains the correlation between predictive performance and power, as shown in \cref{fig:knockoffs_benchmarks_r2}, and achieves similar power to the HRT. This suggests (at least for this simulation) the low power of lasso coefficient knockoffs is likely due to the choice of test statistic. However, unlike the HRT, the method is unable to control FDR in all predictive models. In addition to the theoretical issues with dependency between test statistics, knockoffs also lack a way to calibrate of the complete conditional, which may contribute to the FDP inflations in this case. Finally, it is still possible that the power discrepancy between the lasso coefficient statistic and empirical risk is specific to our simulations; establishing the relative power of empirical risk to other statistics is future work.

\end{document}